\documentclass[sigconf,10pt]{acmart}

\renewcommand\footnotetextcopyrightpermission[1]{} 
\makeatletter
\renewcommand\@formatdoi[1]{\ignorespaces}
\makeatother

\usepackage[utf8]{inputenc}

\usepackage{tikz}
\usepackage{amsmath}

\usepackage{caption}
\usepackage{subcaption}
\usepackage{balance}
\usepackage[linesnumbered,ruled,vlined]{algorithm2e}
\usepackage{bm}
\usepackage{mathtools, cuted}

\usepackage{pifont}
\newcommand{\cmark}{\ding{51}}
\newcommand{\xmark}{\ding{55}}

\AtBeginDocument{
	\providecommand\BibTeX{{
			\normalfont B\kern-0.5em{\scshape i\kern-0.25em b}\kern-0.8em\TeX}}}

\copyrightyear{2021} 
\acmYear{2021} 
\acmBooktitle{}
\setcopyright{acmcopyright}\acmConference[MobiCom '21]{The 27th Annual International Conference on Mobile Computing and Networking}{Oct 25-29, 2021}{New Orleans, United States}

\begin{document}

	\title[Seirios]{(Preprint) Seirios: Leveraging Multiple Channels for LoRaWAN Indoor and Outdoor Localization}
	
	\author{Jun Liu, Jiayao Gao, Sanjay Jha, Wen Hu}
	\affiliation{
		\institution{University of New South Wales}
		\city{Sydney}
		\state{NSW}
		\country{Australia}
		\postcode{2052}
	}

	\renewcommand{\shortauthors}{Liu, et al.}

	\begin{abstract}
		
		Localization is important for a large number of
		Internet of Things (IoT) endpoint devices connected by LoRaWAN. 
		Due to the bandwidth limitations of LoRaWAN, existing localization methods without specialized hardware (e.g., GPS) produce poor performance. 
		To increase the localization accuracy, we propose a super-resolution localization method, called \textit{Seirios}, 
		which features a novel algorithm to synchronize multiple \textbf{non-overlapped} communication channels by exploiting the unique features of the radio physical layer to increase the overall bandwidth. By exploiting both the original and the \textbf{conjugate} of the physical layer, \textit{Seirios} can resolve the direct path from multiple reflectors in both \textbf{indoor} and outdoor environments.
		We design a \textit{Seirios} prototype and evaluate its performance in an outdoor area of 100 m $\times$ 60 m, and an indoor area of 25 m $\times$ 15 m, which shows that \textit{Seirios} can achieve a median error of 4.4 m outdoors (80\% samples < 6.4 m), and 2.4 m indoors (80\% samples < 6.1 m), respectively.
		The results show that \textit{Seirios} produces 42\% less localization error than the baseline approach. 
		Our evaluation also shows that, different to previous studies in Wi-Fi localization systems that have wider bandwidth, time-of-fight (ToF) estimation is less effective for LoRaWAN localization systems with narrowband radio signals.

	\end{abstract}

	\maketitle

	\section{Introduction}
	\label{section:introduction}
	
	LoRa is an emerging technology that provides wide-range wireless network coverage for low power embedded Internet of Things (IoT) devices~\cite{chiani2019lora,liando2019known,LoRaAlliance2017, zhang2020exploring, xie2020combating}. LoRaWAN, a MAC-layer protocol based on LoRa, is widely used to provide reliable and secure wireless communication~\cite{adelantado2017understanding, de2017lorawan, Dongare2018, LoRaAlliance2017}. 
	As millions of IoT devices are deployed with valuable assets both \textbf{indoors and outdoors}~\cite{liando2019known,peng2018plora,talla2017lora,xu2019measurement}, localization becomes an important service
	that can enable a wide range of location-based applications.
	GPS is a popular technology for acquiring such location information \textbf{outdoors}. However, it requires specialized hardware (i.e., GPS receiver) with additional cost, and it does not
	work indoors.
	
	An alternative is to exploit common localization algorithms with existing infrastructure (i.e., gateways) of LoRaWAN, such as received signal strength (RSS)-based location fingerprinting, and time difference of arrival (TDoA)-based or angle of arrival (AoA)-based triangulation, to locate embedded IoT devices; this requires no extra hardware (e.g., a GPS module) and may work \textbf{indoors}~\cite{islam2019lorain,Iliev2015}.
	However, one major disadvantage of these approaches is their undesirable localization accuracy. 
	Taking LoRaWAN as an example, according to one LoRaWAN geolocation whitepaper,\footnote{
		Hyperlink will be revealed in the final version.
	} RSS-based localization can only provide 1,000-2,000 m accuracy. While TDoA-based localization algorithms claim to provide 20-200 m accuracy, an outdoor evaluation in a public LoRaWAN shows that such algorithms achieved a median accuracy of 200 m only~\cite{podevijn2018tdoa}. 
	Such poor localization accuracy cannot meet the requirements of many applications such as geofencing and asset tracking. 
	
	A major performance bottleneck of radio-based localization is the radio multipath effect. Previous research shows that both RSS- and TDoA-based localization methods suffer from significant errors due to the multipath effect~\cite{Gu2018,Chintalapudi2010,wang2017d}.
	Recently, researchers have proposed many approaches to improve the localization accuracy for LoRa~\cite{Chen2019,Gu2018, wolf2018improved, Nandakumar2018}. 
	However, those approaches require either extra hardware (e.g., drones) or modification of the end nodes. Conversely, super-resolution algorithms~\cite{Kotaru2015, Dynamic-MUSIC,Soltanaghaei2018,Vasisht2016,Xie2019, xiong2013arraytrack, Xiong2015, Joshi2015, Kotaru2018,lifs,Ma2019} that have been well investigated to resolve the multipath effect and improve the accuracy of Wi-Fi-based localization have not yet been studied for LoRaWAN. This inspired us to study  super-resolution algorithms with an existing LoRaWAN infrastructure for better localization performance.
	
	Super-resolution algorithms extract the significant reflectors (though strongly coherent) from incoherent channel state measurements and resolve the direct path. Incoherent channel state measurements can be obtained with antenna arrays~\cite{xiong2013arraytrack} or multiple sub-carriers in wideband signals~\cite{Kotaru2015}.
	Kotaru et al. propose SpotFi to utilize multiple subcarriers as virtual sensors with only three physical antennas, which can reduce the cost of large antenna arrays but provide accurate localization with Wi-Fi signals.
	However, the LoRaWAN narrowband signal has only one carrier, and the radio signals in different communication channels are
	naturally \emph{out of sync}, making it difficult to utilize multiple carriers as virtual antennas. Chronos~\cite{Vasisht2016} proposes synchronizing multiple Wi-Fi channels through two-way channel state information (CSI) measurements and packet exchange. However, this approach is difficult to apply in LoRaWAN since its
	data rates can be very slow (e.g., 300 bps), making two-way CSI communication costly; to the best of our knowledge, embedded LoRaWAN radio transceivers cannot measure CSI. ToneTrack~\cite{Xiong2015} proposes utilizing the phase slope of the CSIs of the subcarriers to synchronize overlapped wideband Wi-Fi channels. However, LoRaWAN channels are not overlapped, and the phase slope of a narrowband is not as clear as a wideband such as Wi-Fi. This makes it difficult to synchronize LoRaWAN non-overlapped narrowband channels using this method directly.
	Further, even considering all eight channels (a LoRaWAN device can support up to eight channels with frequency hopping), with the time-of-flight (ToF) estimation method in SpotFi, the overall 1.6 MHz LoRaWAN bandwidth still has a poor resolution of 625 $ns$, equating to 125 m of radio propagation and making it challenging to resolve multipaths in clutter environments.
	
	In summary, the two main challenges for LoRaWAN localization using super-resolution algorithms are:
	\begin{itemize}
		\item the unsynchronized channel state measurements in different \textbf{non-overlapped}  communication channels
		\item insufficient resolution for solving multipath with super resolution algorithms.
	\end{itemize}
	
	To this end, we propose \textit{Seirios},\footnote{\textit{Seirios} (\textit{Sirius}) is the ancient Greek god or goddess of the Dog Star, which is the brightest star in the night sky and an important reference for celestial navigation around the Pacific Ocean.} 
	which exploits the CSI of multiple channels as incoherent measurements and utilizes super-resolution algorithms on multiple anchors (i.e. gateways) to provide accurate localization for LoRaWAN devices
	both \textbf{indoors} and \textbf{outdoors}. It works with legacy LoRaWAN devices, and the gateways or access points (APs) can be acquired with off-the-shelf hardware, 
	making it cost-effective to deploy. 
	
	\textit{Seirios} exploits the unique structure of the radio signals (i.e., linear chirps that sweep the whole band; see Sec.~\ref{design:lorapremier} for details) to synchronize them in multiple non-overlapped LoRaWAN
	channels without two-way communications. 
	In addition to the channel state measurements themselves, \textit{Seirios} utilizes the conjugate of the measurements, which doubles the total amount of 
	information, for better multipath resolution. The contributions of this paper are as follows.

	\begin{itemize}
		
		\item We propose a novel interchannel synchronization algorithm to obtain the synchronized CSI of non-overlapped multiple channels by exploiting the unique structure of the LoRaWAN physical layer. Compared to prior work~\cite{Vasisht2016}, our approach \textbf{does not require two-way communications in CSI measurements}, which is more applicable to LoRaWAN architecture.

		\item We propose doubling the amount of channel information by utilizing both \textbf{the original and the conjugate} of the CSI to increase the numbers of multipaths that the super-resolution algorithms can resolve (up to six reflectors for our prototype AP implementation with two antennas), thus improving the accuracy of localization. 
		
		\item We design and implement a prototype of \textit{Seirios} with software-defined radios (SDRs) as the APs and off-the-shelf embedded LoRaWAN devices. Our evaluation in a 100 m $\times$ 60 m outdoor area
		and a 25 m $\times$ 15 m indoor area shows that \textit{Seirios} achieves a median localization error of 4.4 m and 2.4 m respectively, which are approximately two times smaller than the baseline approaches.
		Different to observations in previous studies with Wi-Fi localization systems, our results show that ToF estimation is less effective for narrowband LoRaWAN localization.
	\end{itemize}

	\section{Related Work}
	\label{section:relatedwork}

	\subsection{LoRa Localization}
	
	Localization systems with, for instance, Wi-Fi, Bluetooth Low Energy (BLE) and cellular network etc. have achieved sub-meter level accuracy as shown in Table~\ref{table:comparisonofrelatedwork}. However, those technologies cannot provide lower power and long range communications for IoT applications. With the rapidly increasing popularity of LPWAN technologies for lower power and long range communications and the importance of localization, recent research has attempted to address the challenges and error sources based on LPWAN infrastructure to 
	improve localization accuracy~\cite{li2020location}.
	Rajalakshmi et al. propose a multi-band backscatter system for localization with subcentimeter-sized devices~\cite{Nandakumar2018}, which is, however, not compatible with legacy devices.
	Chen et al. propose an amplitude-based anti-multipath method using a LoRaWAN radio signal that achieves 4.6 m accuracy indoors~\cite{Chen2019}. However, this approach requires specially designed antennas as opposed to the common omnidirectional antennas. Besides, mounting a LoRaWAN receiver in a flying drone to collect the radio signal from a transmitter at multiple locations is not suitable for conventional
	stationary LoRaWAN gateway deployment.
	To this end, we propose \textit{Seirios} to exploit the existing infrastructure to locate the legacy devices with stationary gateways.

	\begin{table}[htb]
		
		\caption{Comparison of related work}
		\label{table:comparisonofrelatedwork}
		\centering
		\small{
			\begin{tabular}{ccccc}
				\toprule
				\multicolumn{1}{c}{Research} &
				\multicolumn{1}{c}{Technology} &
				\multicolumn{1}{c}{Low Power} &
				\multicolumn{1}{c}{Range} &
				\multicolumn{1}{c}{Accuracy} \\ 
				\midrule
				\cite{Xiong2015,Kotaru2015,Soltanaghaei2018,Vasisht2016} & Wi-Fi & - & 12-25 m & <0.9 m\\
				\cite{hou2017monte,lazik2015alps,bargh2008indoor} & Bluetooth & \cmark & 10 m & $\approx$ 1 m \\
				\cite{kumar2014lte,sun2005signal} & Cellular & \xmark & 35-60 m & $\approx$ 0.85 m\\
				\cite{Vasisht2018,Nandakumar2018,ma2017minding} & Backscatter & \cmark & <10 m & <0.5m \\
				\cite{aernouts2020tdaoa,podevijn2018tdoa} & LPWAN & \cmark & 500 m+ &  >100 m\\
				OwLL \cite{bansal2021owll} & LPWAN & \cmark & 500 m+ &  $\approx$ 9 m\\			Seirios (outdoor) & LPWAN & \cmark & 100 m  & $\approx$ 5 m \\
				WideSee \cite{Chen2019} & LPWAN & \cmark & 40 m & 4.6 m\\
				Seirios (indoor) & LPWAN & \cmark & 25 m & 2.4 m \\
				\bottomrule
			\end{tabular}
		}
	\end{table}

	\subsection{Channel Combination}
	\label{sec:relatedwork:channelcombination}
	Increasing the bandwidth is a useful approach to increase the localization accuracy. Xiong et al. propose ToneTrack~\cite{Xiong2015} to utilize a channel-combining algorithm to increase the bandwidth for finer radio multipath resolution. However, this approach is for overlapped wideband (Wi-Fi) signals only. Nevertheless, it inspired us to combine non-overlapped narrowband LoRaWAN signals to increase the bandwidth for localization (Sec.~\ref{section:sync}). As the resolution of the combined bandwidth signal is still poor (i.e., 125 m), it cannot be used in localization directly. Bansal et al. propose OwLL~\cite{bansal2021owll} to exploit TV whitespace band (up to hundreds of MHz) to increase the accuracy of LoRaWAN localization. There are two major differences between OwLL and \textit{Seirios}. First, OwLL sends hundreds of packets to cover up to tens or hundreds of MHz whitespace radio frequencies, while Seirios uses LoRaWAN bandwidth (i.e., 1.6 MHz with eight channels) only. 
	Thus, there are energy consumption implications of transmitting such a large number (e.g., 80 to 120 packets) of packets for localization. In comparison, \textit{Seirios} transmits eight packets only, which is a fraction, i.e., 1/15 to 1/10, as that of OwLL. For limited bandwidth (e.g., 1.6 MHz) scenarios, we will argue that ToF related algorithms are in fact less effective in Sec.~\ref{section:tof} for LoRaWAN localization, and thus \textit{Seirios} is based AoA only. Second, for synchronization, OwLL follows Chime~\cite{gadre2020frequency} to use an extra transmitter to synchronize the phase of multiple base stations (i.e. gateways), while \textit{Seirios} exploits the microstructure of chirps to synchronize the phases of the LoRa packets in multiple channels instead of synchronizing base stations. Please see Table.~\ref{table:comparisonofowllseirios} for the detailed comparison between
	OwLL and \textit{Seirios}.
	
	\begin{table}[htb]
		\caption{Comparison of OwLL~\cite{bansal2021owll} and \textit{Seirios}}
		\label{table:comparisonofowllseirios}
		\centering
		\small{
			\begin{tabular}{lcc}
				\toprule
				\multicolumn{1}{c}{Features} &
				\multicolumn{1}{c}{OwLL \cite{bansal2021owll}} &
				\multicolumn{1}{c}{Seirios}\\ 
				\midrule
				Bandwidth & 400 MHz & 1.6 MHz \\
				Localization time & 20.97s & 0.24s  \\
				Localization technique & TDoA & AoA  \\
				Packets per localization & 80-120 & 8 \\
				\begin{tabular}{@{}l@{}}Battery life \\ (request twice a day)\end{tabular}   & 1-1.8 years & 10+ years  \\
				Synchronization & Base stations & Multiple channels  \\
				Range & 500 m & 100 m \\
				Accuracy & $ \approx 9$ m &  $ \approx 5$ m (outdoors) \\
				\bottomrule
			\end{tabular}
		}
	\end{table}
	
	\subsection{Virtual Antennas}
	Kotaru et al. propose SpotFi to create a virtual sensor array with the number of elements greater than the number of multipaths, thus, overcoming the constraint posed by a limited number of antennas~\cite{Kotaru2015}. That said, the model proposed in SpotFi is designed for ToF-AoA joint estimation with MUSIC, which produces poor accuracy for LoRaWAN. Therefore, we propose a novel model for accurate AoA estimation with ESPRIT to avoid the unreliable ToF estimation. Moreover, we propose utilizing the conjugates of the channel measurements to further increase the number of virtual antennas, which can further improve localization accuracy.

	\subsection{Spatial Smoothing}
	\label{sec:relatedwork:spatialsmoothing}
	
	Spatial smoothing scheme is proposed by Evan et al. to solve coherent signal classification~\cite{evans1981high}. ArrayTrack~\cite{xiong2013arraytrack} is a uniform linear array (ULA) with eight antennas that utilizes spatial smoothing by averaging two adjacent antennas to resolve multipaths to improve the accuracy of AoA estimation. Theoretical studies by Pillai et al.~\cite{pillai1989forward} and Pan et al.~\cite{pan2020enhanced} show that using both forward and conjugated backward spatial smoothing can further improve the number of coherent signals that can be resolved. However, this method does not work for ULAs with a small number of antennas  (e.g., two or three) only, which are available in low-cost hardware. To this end, SpotFi~\cite{Kotaru2015} proposes using a special Wi-Fi signal model with multiple channels for spatial smoothing with three antennas. We note that SpotFi does not make use of the conjugate information due to a significantly larger number of communication channels available in Wi-Fi (e.g., 30 against eight in LoRaWAN). Therefore, the bandwidth constraint problem is unique to LoRa signal studied in this paper. \textit{Seirios} is inspired by both SpotFi and conjugated backward techniques to propose a novel LoRa signal model, which solves coherent multipath signals with a small number of antennas (i.e., two)  and a limited number of channels (i.e., eight).

	\textbf{Summary}:

	\textit{Seirios} is inspired by the channel combination technique of ToneTrack, the virtual antenna model of SpotFi and conjugated backward spatial smoothing. First, it features a channel combination method for non-overlapped narrowband signals to increase the overall bandwidth. Second, instead of using the MUSIC model in SpotFi, \textit{Seirios} utilizes a novel model with virtual antennas to accurately estimate AoA with the ESPRIT algorithm to avoid unreliable ToF estimation. Moreover, it exploits the conjugates of the measurements to further increase the number of virtual antennas and the capacities for multipath resolution.
	
	\section{Background}

	\subsection{LoRa Premier}
	\label{design:lorapremier}

	\begin{table}[htb]
		\caption{Summary of mathematical symbols used.}
		\label{table:terminology}
		\centering
		\small{
			\begin{tabular}{cl}
				\toprule
				\multicolumn{1}{l}{Symbol} & \multicolumn{1}{l}{Definition} \\ 
				\midrule
				SF & LoRa spreading factor\\
				BW & LoRa bandwidth \\
				$F_s$ & Sampling rate  \\
				$T$ & LoRa symbol duration  \\
				$M$ & Number of channels \\
				$P$ & Number of multipaths \\
				$H$ & Channel response \\
				$\bm{R}$ & Covariance matrix \\
				$\bm{A}$ & Steering matrix \\
				$\bm{X}$ & CSI matrix \\
				$\bm{\Phi}$ & Diagonal matrix of the angle of arrival\\
				$\bm{\Gamma}$ & Matrix of path attenuation \\
				$\lambda$ & Chirp rate in linear chirps\\
				$c$ & Speed of light\\
				$d$ & Sensor spacing (i.e., distance between two antennas) \\
				$f_c$ & Carrier frequency\\
				$f_\delta $ & Channel spacing (i.e., frequency between two channel)\\
				\bottomrule
			\end{tabular}
		}
	\end{table}

\begin{figure}[hbt]	
		\centering
		\includegraphics[width=0.75\linewidth]{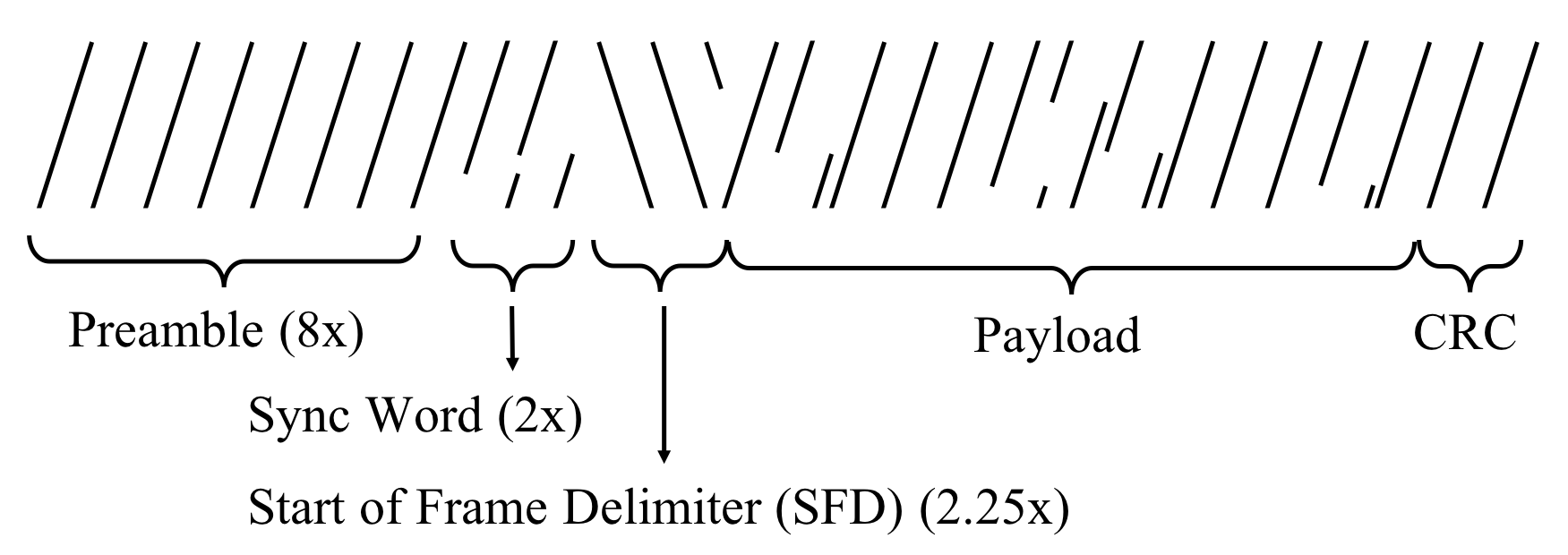}
		\caption{An example physical-layer CSS of the LoRa packet in the time-frequency domain. }
		\label{fig:lora-phy}
		\raggedright Note: The x-axis reflects time, and the y-axis reflects frequency
	\end{figure}

	Before discussing LoRa, note Table~\ref{table:terminology}, which summarizes the mathematical symbols used in the following sections.
	LoRa is modulated with a chirp spread spectrum (CSS)~\cite{chiani2019lora,adelantado2017understanding}. 
	This is configured by the spreading factor ($SF$) and bandwidth ($BW$). $SF$ is defined as an integer from 7 to 12, representing the number of encoded bits per chirp symbol, and $BW$ is the bandwidth of a channel, typically 125 kHz or 500 kHz~\cite{LoRaAlliance2017}. An up-chirp has its frequency increasing linearly, while a down-chirp is the opposite. A chirp is the minimum unit of a LoRa radio signal, and a LoRa packet is modulated as the concatenation of different chirps. The structure of a LoRa packet 
	can be found in Fig.~\ref{fig:lora-phy}.
	
	The preambles are identical to facilitate the packet detection. Preambles consist of a predefined number (e.g., eight for LoRaWAN) of up-chirps, and 
	the frequency of an up-chirp is defined as
	\begin{equation}
		f(t) = \lambda t- \frac{BW}{2}, \quad t \in [0,T), 
		\label{eq:lora:freq}
	\end{equation}
	
	where $\lambda = \frac{BW^2}{2^{SF}}$ is the chirp rate, and $T = \frac{2^{SF}}{BW}$ is the duration of the chirp.
	The phase of up-chirp $\varphi(t)$ can be obtained by integrating $f(t)$ as,
	\begin{equation}
		\varphi(t) = 2\pi\int_{0}^{t}f(\tau)d\tau = 2\pi(\frac{\lambda}{2} t^2- \frac{BW}{2} t),  \quad t \in [0,T). 
		\label{eq:lora:phase}
	\end{equation}
	
	Then, an up-chirp with normalized magnitude can be represented as,
	\begin{equation}
		u(t) = e^{ j \, \varphi(t)},  \quad t \in [0,T). 
		\label{eq:lora:upchrip}
	\end{equation}
	
	Eq.~(\ref{eq:lora:freq}) shows that LoRa CSS modulation utilizes the entire bandwidth, \textbf{making it possible to measure the channel state of the whole band with high resolution by comparing the received chirps with the up-chirp reference, which can facilitate interchannel synchronization} (Sec.~\ref{section:sync}). 
	
	\subsection{Signal Model}
	
	\label{section:signal:model}
	
	\begin{figure}	
		\centering
		\includegraphics[width=0.60\linewidth]{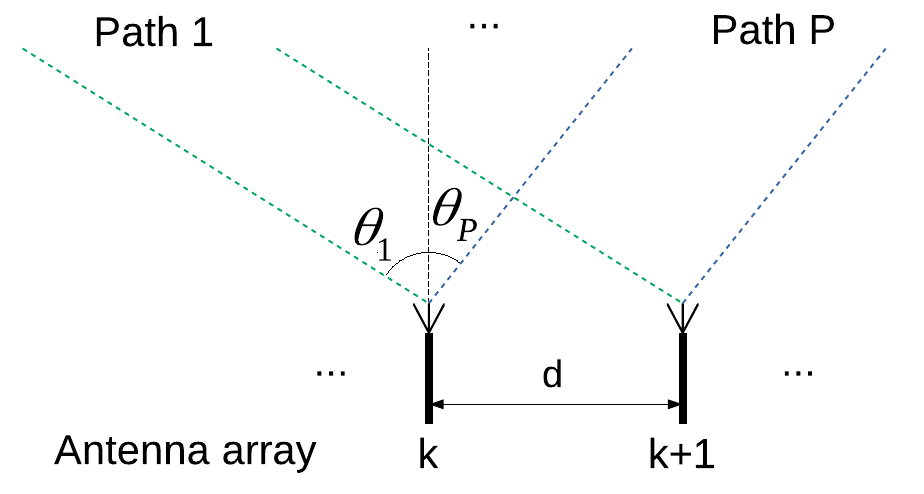}
		\caption{A pair of adjacent antennas in uniform linear array. The AoA is defined as the angle between an incident signal and the array's normal. Antenna spacing $d$ is the distance between adjacent antennas. The source is far from the array; thus, the incident signals of the same path are parallel.}
		\label{fig:AoAdef}
	\end{figure}
	
	Suppose there are $P$ significant paths between a sender and a receiver. Since the signal is narrowband, the channel response of each path can be modeled as a complex value $\alpha_p (p=1 \dots P)$, representing the amplitude attenuation and phase shift compared to the original signal $u(t)$ that has been sent. 
	The received signal $r(t)$ is the sum of multipath replicas of the original signal $u(t)$ as
	\begin{equation}
		r(t) = \sum_{p=1}^P \alpha_p u(t-\tau_p),
		\label{eq:r(t)}
	\end{equation}
	
	where $\tau_p$ is the ToF of the $p$-th path. With a Fourier transform, we can transfer Eq~(\ref{eq:r(t)}) into the frequency domain as,
	\begin{equation}
		R(f) = U(f) \sum_{p=1}^P \alpha_p e^{-j2\pi  f \tau_p},
		\label{eq:R(f)}
	\end{equation}
	
	where $f$ is the frequency. Therefore, we can obtain the channel response by
	\begin{equation}
		H(f) = \frac{R(f)}{U(f)}= \sum_{p=1}^P \alpha_p e^{-j2\pi  f \tau_p },
		\label{eq:H(f)}
	\end{equation}
	
	Suppose we have $M$ equally spaced channels with frequency spacing $f_\delta$ and $K$ antennas.
	We use the central frequency to represent the frequency of the whole narrowband channel and $H_{k,i}$ to represent the channel response of the $i$-th ($i=1 \dots M$) channel measured by the $k$-th ($k=1 \dots K$) antenna.
	Each path has an AoA $\theta_p (p=1 \dots P)$, as depicted in Fig.~\ref{fig:AoAdef}. Therefore, $H_{k,i}$ can be represented by
	\begin{align}
		H_{k,i} = \sum_{p=1}^P \alpha_p e^{-j2\pi [f_c+(i-1) \cdot f_\delta] \tau_p } e^{-j (k-1) 2\pi d sin(\theta_p) f_c /c} ,
		\label{eq:Hki}
	\end{align}
	
	where $d$ is the antenna spacing---that is, the distance between the adjacent antennas ($k$ and $k+1$)---and $c$ is the speed of light. 
	For simplicity, we use $\Phi(\theta_p)$ or $\Phi_p$ to represent the phase shift caused by the AoA $\theta_p$ and  $\Omega(\tau_p)$ or $\Omega_p$ caused by 
	the ToF $\tau_p$. We have
	\begin{align}
		\Phi_p &= \Phi(\theta_p) = e^{-j2\pi d sin(\theta_p) f_c /c }\label{eq:phi:},\\
		\Omega_p &= \Omega(\tau_p) = e^{-j2\pi  f_\delta \tau_p },\\
		\gamma_p &= \alpha_p e^{-j2\pi f_c \tau_p}, \label{eq:model:gamma}\\
		H_{k,i} &= \sum_{p=1}^P \gamma_p \Phi_p^{k-1} \Omega_p^{i-1}.
	\end{align}

	For linear chirps, as suggested by Eq.~(\ref{eq:lora:freq}), frequency increases linearly and monotonically with time. With Eq.~(\ref{eq:H(f)}), the channel response for continuous frequency can be measured with linear chirps. For example, 125 kHz LoRa chirp with central frequency 920 MHz can be used to measure the channel response from 919.9375 MHz to 920.0625 MHz. \textit{Seirios} utilizes this microstructure of LoRa signals to obtain the channel response of continuous frequency and perform interchannel synchronization (see Sec.~\ref{section:sync} for details). \subsection{MUSIC and ESPRIT}
	
	\label{section:background:musicandesprit}
	
	MUSIC~\cite{schmidt1986multiple,Kotaru2015, Dynamic-MUSIC} and ESPRIT~\cite{roy1986esprit, roy1989esprit} are two popular super-resolution algorithms, and have been shown to resolve the multipath effect in wideband radio (e.g., Wi-Fi) localization systems.
	
	Previous research has shown that increasing the number of antennas (e.g., up to eight) can improve the localization accuracy~\cite{xiong2013arraytrack} at the cost of extra hardware (antenna). To this end, \textit{Seirios} can operate with two antennas, which is cost-effective.  For a pair of antennas ($k$ and $k+1$) with $M$ LoRaWAN channels, the measurements matrix $\bm{X}_{MU}$ for MUSIC can be organized as:
	\begin{equation}
		\bm{X}_{MU} = [H_{k,1} \cdots H_{k,M},H_{k+1,1} \cdots H_{k+1,M}]^T.
	\end{equation}
	
	The steering vector required by MUSIC is
	\begin{equation}
		\bm{\vec{a}} = \begin{bmatrix}
			1 
			\cdots 
			(\Omega(\tau))^{M-1} ,
			\Phi(\theta) 
			\cdots 
			\Phi(\theta)(\Omega(\tau))^{M-1}
		\end{bmatrix}^T.
		\label{eq:steeringvector}
	\end{equation}
	
	We can then utilize the classical MUSIC algorithm with $\bm{X}_{MU}$ and $\bm{\vec{a}}$ to estimate each radio signal path
	between a sender and a receiver by searching the AoA and ToF that can generate peaks on the MUSIC pseudo-spectrum.
	
	Alternatively, we can also use ESPRIT to solve multipaths. The measurement matrices $\bm{X}_{ES,k}$ and $\bm{X}_{ES,k+1}$ for antenna $k$ and $k+1$ can be organized as
	\begin{align}
		\bm{X}_{ES,k} &= [H_{k,1} \cdots H_{k,M}]^T, \label{eq:esprit:orig:Xk}\\
		\bm{X}_{ES,k+1} &= [H_{k+1,1} \cdots H_{k+1,M}]^T. \label{eq:esprit:orig:Xk+1} 
	\end{align}
	
	We can compose a model for ESPRIT as 
	\begin{align}
		\begin{bmatrix}
			\bm{X}_{ES,k}\\
			\bm{X}_{ES,k+1}\\
		\end{bmatrix}
		&=\begin{bmatrix}
			\bm{A} \\
			\bm{A}\bm{\Phi} \\
		\end{bmatrix}
		\bm{\Gamma} + \bm{\epsilon}, \label{eq:signalmodel}
	\end{align}
	
	where 
	$\bm{\Phi}=diag(\Phi_1 \ldots \Phi_P)$, $\bm{\Gamma}= [\gamma_1 \ldots \gamma_P]^T$, $\bm{\epsilon}$ is the noise, and $\bm{A}$ is a steering matrix. By solving $\bm{\Phi}$ with ESPRIT, the AoAs are estimated.
	
	We skip the details of MUSIC and ESPRIT for brevity.

	\section{Design}
	\label{section:design}

	As discussed in Sec.~\ref{section:introduction}, \textit{Seirios} utilizes triangulation with multiple APs. The key to improving localization accuracy is to improve AoA estimation. In this section, we will focus on interchannel synchronization (Sec.~\ref{section:sync}) and improved super-resolution algorithms (Sec.~\ref{section:esprit}) to overcome the challenges discussed in Sec.~\ref{section:introduction}. This will subsequently improve the accuracy of AoA estimation as well as the performance of the localization system. 
	
	\subsection{System Design}
	\label{section:systemdesign}
	
	\begin{figure*}[htb]
		\centering
		\includegraphics[width=0.80\linewidth]{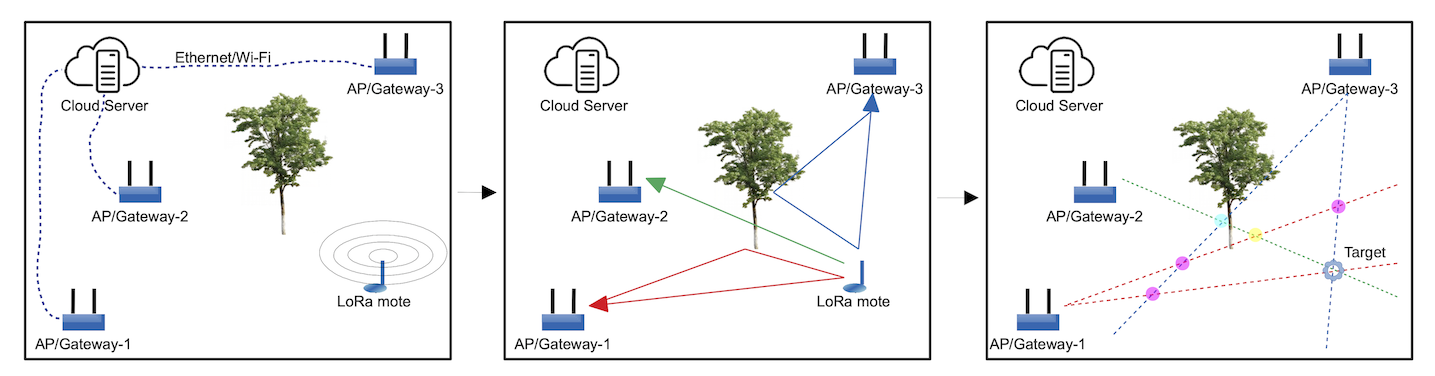}
		\caption{An example of \textit{Seirios} localization system. The locations of gateways are known in advance. In the cloud server, the system estimates the location of the transmitter via maximum likelihood gateway fusion introduced in Sec.~\ref{subsec:fusion}.}
		\label{fig:architecture}
	\end{figure*}
	
	\textit{Seirios} is designed to exploit multiple LoRaWAN APs to locate a transmitter. The location and antenna direction of APs are known in advance and stored in the cloud. Here, the APs sense the radio signal transmitted from the transmitter before relaying them to the cloud server. We assume that there is only one transmitter and no concurrent transmission at the same channel. Then, the transmitter is located, after which the cloud server synchronizes the channels and performs an AoA estimation from the radio signal measured at each AP. Each AP in \textit{Seirios} has at least two synchronized antennas. The distance between the two antennas is slightly less than half of the wavelength. If there is one path only between the transmitter and an AP, the AoA can be calculated directly by comparing the phase difference of the received signal between two antennas. However, in reality, there are multiple paths due to radio signal reflection. Even though the line-of-sight (LoS) path exists, other paths may cause significant errors in the AoA estimation due to radio signal self-interference.

	Prior studies show that the number of significant reflectors in an indoor environment is 5.05 on average, with a standard deviation of 1.95~\cite{Vasisht2016}. 
	For an outdoor or uncluttered indoor environment, there will be 
	even fewer significant reflectors (e.g., four). Therefore,
	one of the research questions concerns how to accurately estimate the AoA of a limited number of paths in such
	environments (both indoor and outdoor)
	with narrowband radio signals (e.g., LoRaWAN).
	
	Fig.~\ref{fig:architecture} shows an illustrated example where \textit{Seirios} decomposes the radio wave multipaths generated 
	by the tree reflectors, estimates the AoA of the radio wave in different APs, and locates the target successfully.

	\subsection{Channel State}
	\label{section:csi}
	
	Channel state information (CSI) represents how signals at certain carrier frequencies propagate from the transmitter to the receiver along multiple paths~\cite{Ma2019}. It has been widely used in Wi-Fi signal based localization systems \cite{Kotaru2015,Xiong2015,Soltanaghaei2018,lifs,Xie2019}. To measure CSI, a Wi-Fi transmitter sends packets whose preamble contains pre-defined training symbols for each subcarrier. When the training symbols are received, the receiver can measure CSI by comparing the amplitude and phase of the received symbols with the pre-defined training symbols. 
	
	For LoRa, similar CSI can be obtained. At the receiver, the radio signal can be sampled as a complex sequence of $I$ and $Q$ components. \textit{Seirios} detects the preamble and applies the digital processing algorithms introduced in~\cite{ghanaatian2019lora} for precise carrier frequency offset (CFO) and sampling time offset (STO) calibration. In the following discussion, we assume that all LoRa chirps are well calibrated. Since both the sender and the receiver know the preamble, this can be regarded as a training sequence. Here, LoRa CSI can be obtained by comparing the received preambles with the pre-defined preambles (i.e., linear up-chirps).

	Furthermore, we can sum up the repeating up-chirps in the preamble to improve the signal-to-noise ratio (SNR) in CSI estimation. There are two factors ensuring that the up-chirps can be summed up as follows. Firstly, since the frequency modulation for up-chirp is symmetric (see Eq.~(\ref{eq:lora:freq})), the phase will roll back to its initial state after the period of one chirp, and, thus, \textbf{all these up-chirps have the same phase}. Moreover, the preamble only lasts for a small amount of time and the channel response does not change during this period. Thus, \textbf{all these up-chirps have the same CSI}. Therefore, we can combine all the up-chirps by
	\begin{equation}
		\bar{r}(t) = \frac{1}{N_{Preamble}} \sum_{l=1}^{N_{Preamble}} r^{(l)}(t),
	\end{equation}
	where $N_{Preamble}$ stands for the number of up-chirps in the preamble, 
	$r^{(l)}(t)$ is the $l$-th up-chirp received, and $\bar{r}(t)$ represents the summation of received up-chirps. 
	
	To this end, LoRa CSI can be measured by,
	\begin{equation}
		CSI = \frac{1}{T} \int_0^T \bar{r}(t) \cdot u^*(t) dt, 
		\label{eq:CSIa}
	\end{equation}
	
	where $u(t)$ is the zero-phased up-chirp defined in Eq.~(\ref{eq:lora:upchrip}), and $(.)^*$ denotes the conjugate transpose. Eq.~(\ref{eq:CSIa}) is similar to the pulse compression technique, which is used in LoRa demodulation to significantly increase the SNR.
	
	Different to Chronos~\cite{Vasisht2016}, which estimates the CSI of a wireless link by comparing the CSI measurements in the two end nodes of the link, \textit{Seirios} estimates the CSI by comparing the received up-chirps in the preamble with the reference on the receivers only. 
	The advantage of our approach is two-fold. First, 
	none of the embedded LoRaWAN devices (transmitters) is capable of measuring CSI. \textit{Seirios} requires CSI measurements
	in one end (i.e., LoRaWAN gateway) only, while Chronos requires both ends to measure CSI. This makes  \textit{Seirios} cost-effective to deploy because the number of embedded devices is orders of magnitude
	more than that of gateways. Second, the data rates of LoRaWAN (can be as low as 300 bps) are orders of magnitude
	smaller than those of Wi-Fi (the lowest is 1 Mbps). Therefore, transmitting the CSI measurements between
	two ends incurs significantly more time and (energy) costs in LoRaWAN. 
	
	We have discussed the method for LoRa CSI measurements above. Below, we will discuss how CSI can be used in localization.
	
	Recall that both MUSIC and ESPRIT utilize channel response to resolve radio signal multipaths (see Sec.~\ref{section:signal:model} and Sec.~\ref{section:background:musicandesprit}). The main difference between CSI and channel response is a random phase shift that is introduced in CSI because the transmitter and the receiver are not synchronized. For wideband Wi-Fi signals, CSIs for each subcarrier are measured at the same time. The phase shift is the same for each subcarrier, so it does not affect the results of super-resolution algorithms. Therefore, CSI can be used directly by the super-resolution algorithms.
	
	However, for the $i$-th narrowband channel of LoRa measured by antenna $k$, $CSI_{k,i}$ is phase-shifted from $H_{k,i}$ (see Eq.~(\ref{eq:Hki})) as
	\begin{equation}
		CSI_{k,i} = H_{k,i} \cdot e^{j2\pi(\phi^{tx}_{i} - \phi^{rx}_{i})}, 
		\label{eq:csi=h*phaseshift}
	\end{equation}
	where $\phi^{tx}_{i}$ and $\phi^{rx}_{i}$ are the initial phases of the transmitter and the receiver, respectively. Since a LoRa channel (here LoRa channels may be viewed as the Wi-Fi subcarriers) is measured individually, the random phase shift is different in each channel. Therefore, unlike Wi-Fi CSI, LoRa CSI must be calibrated before it can be used by the super-resolution algorithms.
	
	Calibrating the phase shift is equivalent to synchronizing the LoRa channels by solving $\phi^{tx}_{i} - \phi^{rx}_{i}$ in Eq.~(\ref{eq:csi=h*phaseshift}).
	However, solving the synchronization problem between a transmitter and a receiver is challenging. Chronos~\cite{Vasisht2016} proposed measuring CSI on both the transmitter and the receiver to eliminate the phase shift, yet the method requires extra hardware to measure CSI in embedded LoRaWAN transmitters, which is undesirable. 
	
	Conversely, in the context of localization,
	we do not need to estimate the absolute values of the phase shifts. Instead, we need an identical phase shift---say, the phase shift of the first channel---for all calibrated channels to be used by the  super-resolution algorithms. In the next section, we will discuss how to exploit the microstructure of LoRa signal for synchronization.

	\subsection{Interchannel Synchronization}
	\label{section:sync}

	\begin{figure}[htb]
		\centering
		\begin{subfigure}[t]{0.48\linewidth}
			\centering
			\includegraphics[width=\textwidth]{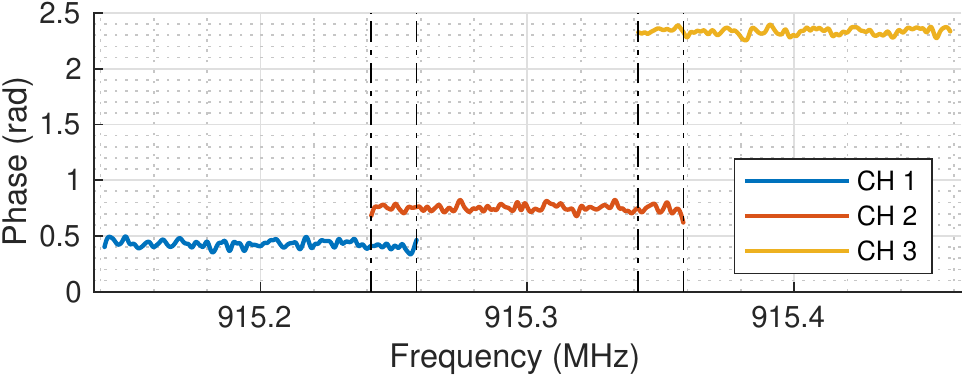}
			\caption{}
			\label{fig:phase:showcase:unsync}
		\end{subfigure}
		\hfill
		\begin{subfigure}[t]{0.48\linewidth}
			\centering
			\includegraphics[width=\textwidth]{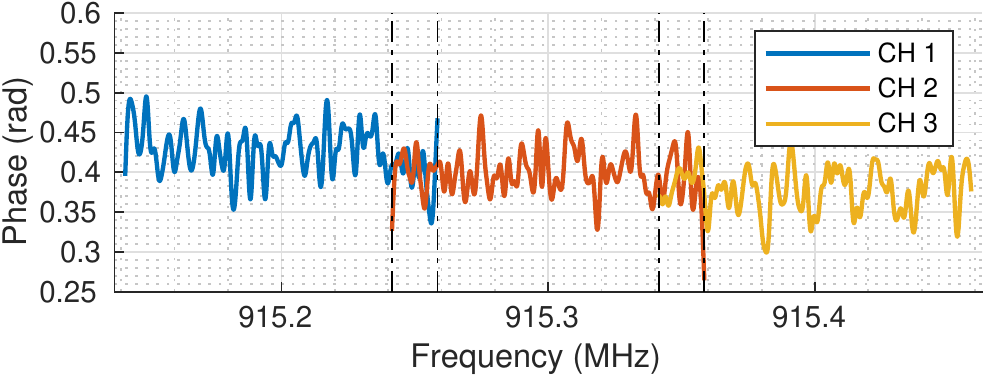}
			\caption{}
			\label{fig:phase:showcase:sync}
		\end{subfigure}
		\begin{subfigure}[t]{0.48\linewidth}
			\centering
			\includegraphics[width=\textwidth]{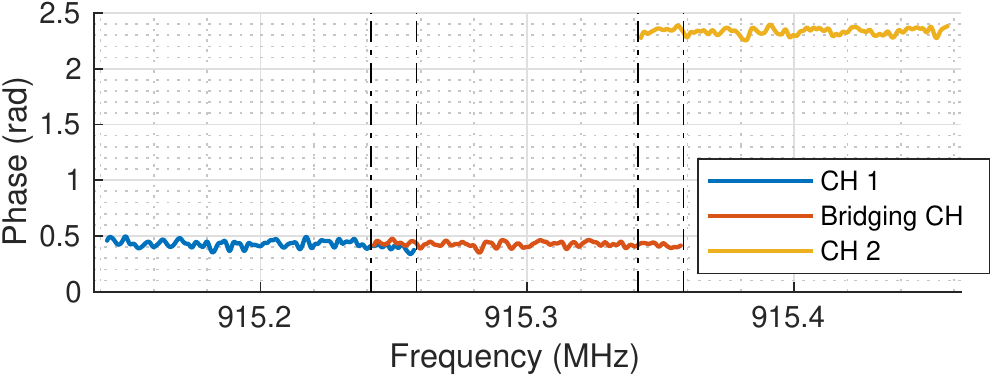}
			\caption{}
			\label{fig:phase:showcase:unsync:nonoverlapped}
		\end{subfigure}
		\hfill
		\begin{subfigure}[t]{0.48\linewidth}
			\centering
			\includegraphics[width=\textwidth]{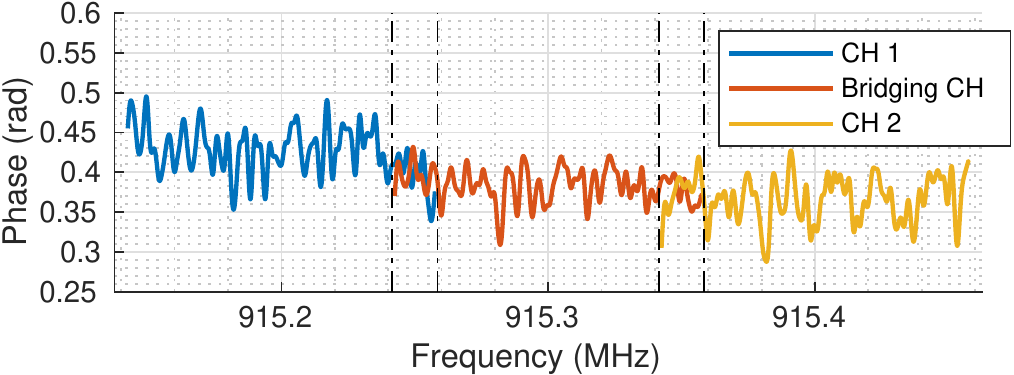}
			\caption{}
			\label{fig:phase:showcase:sync:nonoverlapped}
		\end{subfigure}
		\caption{The phase of channel state $g(f)$ for continuous frequency under different conditions: (a) lack of synchronization between transmitters and receivers introduces random phase shift; (b) interchannel synchronization exploits the overlapped band to eliminate the phase shift; (c) for non-overlapped channels, a (virtual) bridging channel can be generated by averaging the phases in two adjacent (co-phased) channels; (d) use a generated (virtual) bridging channel to synchronize two adjacent channels.}
		\label{fig:phase:showcase}
	\end{figure}

	First of all, let us briefly review the related work discussed in Sec.~\ref{sec:relatedwork:channelcombination}. ToneTrack~\cite{Xiong2015} has proposed a method to align overlapped Wi-Fi channels. (Adjacent channels are regarded as the special forms of the overlapped channels in this paper, to be differentiated from non-overlapped channels.) It first 
	equalizes the phase slope in the frequency domain; then, 
	it aligns the phase of the last subcarrier of the first wideband channel and the first subcarrier of the second wideband channel. With these two steps, the second wideband channel can be concatenated to the first channel to form a wider band. However, this method only works for overlapped channels. By contrast, 
	\textit{Seirios} improves the method to operate on non-overlapped narrowband channels. 
	
	Different to a Wi-Fi signal, which has multiple subcarriers to form a phase slope, a narrowband LoRaWAN signal has only one carrier. To form a similar phase slope for the LoRaWAN signal, we exploit the microstructure of LoRa chirps. For illustration, we define $g(f)$ as the channel state for a continuous frequency range from the lower bound to the upper bound of the bandwidth. Fig.~\ref{fig:phase:showcase:unsync} shows an example of the phase slopes for three channels (915.2 MHz, 915.3 MHz, and 915.4 MHz). Note the random phase shift in between is caused by lack of synchronization (see the discussion in Sec.~\ref{section:csi}).
	Eq.~(\ref{eq:lora:freq}) shows that frequency $f$ and time $t$ have a linear relationship. Therefore, we can estimate $g_i(f)$ for channel $i$ as
	\begin{equation}
		g_i(f) = \bar{r}_i(t(f))u^*(t(f))= \bar{r}_i(\frac{f-f_i+\frac{BW}{2}}{\lambda})u^*(\frac{f-f_i+\frac{BW}{2}}{\lambda}), 
		\label{eq:CSI}
	\end{equation}
	where $f \in [f_i-\frac{BW}{2},f_i+\frac{BW}{2}]$, $f_i$ is the carrier frequency of channel $i$.
	Fig.~\ref{fig:phase:showcase:unsync} depicts the phase of $g_i(f)$ for $i=1,2,3$.
	
	However, unlike the stable phase slope illustrated by \cite{Xiong2015}, the phase slope of the LoRa narrowband channel is full of noise even when the SNR is high (see Fig.~\ref{fig:phase:showcase:sync}, where the SNR is 10 dB). This is because that the phase change in a narrowband (e.g., in channel 1 the phase decreases by approximately 0.05 rad, as shown in Fig.~\ref{fig:phase:showcase:sync}) is much less than that of a Wi-Fi wideband (e.g., approximately 3.0 rad in~\cite{Xiong2015}), and, thus, the narrowband is less robust to the noise than Wi-Fi even with the same noise level. To better estimate the phase offset between two channels, \textit{Seirios} \textbf{averages} the phase offset within the overlapped frequency to reduce the noise. By compensating the phase offset of the second channel to align with the first channel, the second channel can now be concatenated to the end of the first channel, as shown in Fig.~\ref{fig:phase:showcase:sync}.

	For non-overlapped channels, ToneTrack states that estimating the correct amount of phase offset is challenging. To this end, for narrowband radio signals, we propose to generate a (\textbf{virtual}) intermediate channel response as a bridge to assist the synchronization. In the microbenchmark in Sec.~\ref{sec:microbenchmark}, we show that channel response varies slowly with frequency, making it possible to generate the virtual intermediate channel with small errors by averaging the two adjacent channels. Taking Fig.~\ref{fig:phase:showcase:unsync:nonoverlapped} as an example, we can obtain the (virtual) intermediate channel response (shown as a bridging channel in the figure) by averaging the phases of (co-phased) channel 1 and channel 2. Similar to the overlapped channels, non-overlapped channels can now be synchronized with the (virtual) intermediate channel (Fig.~\ref{fig:phase:showcase:sync:nonoverlapped}). 
	
	Unlike Wi-Fi wideband, which measures multiple CSIs in one packet, a LoRa narrowband can only measure one CSI per packet. To measure the CSI of multiple channels, at least one packet on each channel should be transmitted within the coherence time.  
	A LoRaWAN end device has eight channels of 125 kHz with 200 kHz spacing. Eight packets should be transmitted on eight different channels to measure their CSI. LoRaWAN channels are separated by 75 kHz guard bands, and the (virtual) intermediate channel has a 25 kHz overlap with the adjacent channels. In practice, \textit{Seirios} leaves out the first and the last 4 kHz because the quality of phase 
	estimation in transient is poor. Thus, the overlapped frequency of one LoRaWAN channel (with the virtual bridging channel) is 17 kHz (13.6\% of the bandwidth), which is sufficient for synchronization (see our evaluation in Sec.~\ref{section:evaluation:sync} for more details). 
	The synchronized CSI can later be used in super-resolution algorithms for multipath resolution.

	\subsection{Limitation of ToF Estimation}
	\label{section:tof}

	A typical configuration of LoRaWAN has 8 $\times$ 125 kHz channels with the channel spacing of 200 kHz between two adjacent channels. Therefore, we can measure the CSI for each channel with a pair of antennas $k$ and $k+1$ (see Sec.~\ref{section:csi}) as
	\begin{equation}
		\begin{bmatrix}
			x_{k,1} & x_{k,2} & \cdots & x_{k,8}\\
			x_{k+1,1} & x_{k+1,2} & \cdots & x_{k+1,8}
		\end{bmatrix}.
		\label{matrix:csi}
	\end{equation}
	
	Following the spatial smoothing approach used in \cite{Kotaru2015} to maximize the incoherence of measurements for better multipath resolution,
	the measurement matrix for MUSIC is
	\begin{equation}
		\bm{X}_{MUSIC}= 
		\begin{bmatrix}
			x_{k,1} & x_{k,2} & \cdots & x_{k,6}\\
			x_{k,2} & x_{k,3} & \cdots & x_{k,7}\\
			x_{k,3} & x_{k,4} & \cdots & x_{k,8}\\
			x_{k+1,1} & x_{k+1,2} & \cdots & x_{k+1,6}\\
			x_{k+1,2} & x_{k+1,3} & \cdots & x_{k+1,7}\\
			x_{k+1,3} & x_{k+1,4} & \cdots & x_{k+1,8}
		\end{bmatrix}.
	\end{equation}

	We can obtain a steering vector $\vec{\bm{a}}$ according to Eq.~(\ref{eq:steeringvector}) with $M=3$. 
	With $\bm{X}_{MUSIC}$ and the steering vector $\vec{\bm{a}}$, we can use MUSIC to estimate AoA and ToF jointly. The super-resolution algorithm searches all combinations of $\theta$ and $\tau$ for the steering vector and calculates the value of a pseudo-spectrum function to find the peaks where the most likely estimations are located. 
	However, as shown in our evaluation in Sec.~\ref{sec:evaluation}, the localization performance is poor and sometimes even worse than the baseline. 
	Thus, we observe the following:
	\begin{itemize}
		\item ToF resolution is 625 $ns$ (125 m), making the paths with ToF difference indistinguishable. In the pseudo-spectrum, those estimations merge into one peak which is away from the ground truth. Therefore, the attempt to resolve multipaths fails. With a wider bandwidth (e.g., 20 MHz as Wi-Fi), accuracy can be improved.
		\item AoA estimation is sensitive to ToF accuracy. If ToF is estimated poorly, it will be noisy for AoA estimation, making AoA estimation worse than the baseline.
	\end{itemize}
	
	Therefore, we look for other algorithms that can avoid using ToF estimation for direct path resolution.

	\subsection{AoA Estimation with Conjugate}
	\label{section:esprit}

	ESPRIT~\cite{roy1986esprit,roy1989esprit,hu2014esprit} takes advantage of rotational invariance for AoA estimation. However, it is not used widely in Wi-Fi super-resolution-based localization schemes due
	to its inferior performance compared to MUSIC~\cite{oumar2012comparison}. One possible reason for this is that ESPRIT does not estimate ToF as MUSIC does. Since LoRa has orders of magnitude less bandwidth than Wi-Fi (see Sec~\ref{section:systemdesign}), ToF estimation is inaccurate and will reduce the overall performance. With ESPRIT, we can exploit all the information for AoA estimation only, which may have better performance than AoA-ToF joint estimation. 
	Our results in Secs.~\ref{sec:microbenchmark} and~\ref{sec:evaluation} show
	that AoA estimation alone has better performance 
	than AoA-ToF joint estimation.

	The signal model for ESPRIT is shown as Eq.~(\ref{eq:signalmodel}). 
	Measurements are organized in two matrices, $\bm{X}_{k}$ and $\bm{X}_{k+1}$. (We abuse the notations as those in Sec.~\ref{section:tof} for brevity.) A naïve approach is to follow Eq.~(\ref{eq:esprit:orig:Xk}) and (\ref{eq:esprit:orig:Xk+1}) for \textbf{a pair of antennas} to form matrices with 8 $\times$ 1 dimensions. 
	However, their covariance matrix $\bm{R}=E\{\bm{X}_{k}\bm{X}_{k+1}\}$ is rank deficient with rank one, which indicates that the number of multipaths the algorithm can solve is one only. 
	
	As discussed in Sec.~\ref{sec:relatedwork:spatialsmoothing}, spatial smoothing can resolve multipath. However, when averaging subsets of antennas, it also reduces the number of effective antennas. For example, the number of effective antennas in ArrayTrack is reduced from eight to seven after averaging two adjacent antennas~\cite{xiong2013arraytrack}. With two antennas in \textit{Seirios}, there will be only one effective antenna left after spatial smoothing, which can only solve one path. Therefore, we cannot use the spatial smoothing proposed in \cite{xiong2013arraytrack} directly. Instead, \textit{Seirios} follows the technique proposed in SpotFi~\cite{Kotaru2015} to do spatial smoothing with a special signal model to increase the rank to four.
	Our evaluation shows that solving four paths is useful for outdoor localization, but results in poor accuracy for indoor localization. To improve the localization performance in multipath-rich environments, the rank should be further increased.
	
	Pillai et al.~\cite{pillai1989forward} show that conjugated backward spatial smoothing can further improve the number of coherent signals that it can resolve for ULA from $\lfloor K/2 \rfloor$\footnote{$\lfloor * \rfloor$ stands for the integer part of $*$} to $\lfloor 2K/3 \rfloor$ ($K$ stands for the number of antennas). For \textit{Seirios}, $K=2$ and $\lfloor 2K/3 \rfloor = 1$, which means that this technique cannot improve multipath resolution if it is applied directly. However, it shows that the conjugates have extra information and can be utilized in spatial smoothing to further improve multipath resolution. Therefore, we extend the aforementioned signal model to include the conjugates of the measurements to solve more multipaths.

	To this end, we propose to exploit both the original and the conjugate of the measurements. We call this algorithm \textit{conjugated ESPRIT}. $\bm{X}_{k}$ and $\bm{X}_{k+1}$ can be organized with CSIs from Eq.~(\ref{matrix:csi}) 
	\begin{equation}
		\bm{X}_k = \begin{bmatrix}
			x_{k,1} & x_{k,2} & x_{k,3} & x_{k+1,8}^* & x_{k+1,7}^* & x_{k+1,6}^* \\
			x_{k,2} & x_{k,3} & x_{k,4} & x_{k+1,7}^* & x_{k+1,6}^* & x_{k+1,5}^*\\
			\vdots  & \vdots  &  \vdots &   \vdots   &    \vdots  &   \vdots \\
			x_{k,6} & x_{k,7} & x_{k,8} & x_{k+1,3}^* & x_{k+1,2}^* & x_{k+1,1}^* \\
		\end{bmatrix} 
		\label{eq:esprit:Xk}
	\end{equation}
	as a 6 $\times$ 6 matrix, and symmetrically,
	\begin{equation}
		\bm{X}_{k+1} = \begin{bmatrix}
			x_{k+1,1} & x_{k+1,2} & x_{k+1,3} & x_{k,8}^* & x_{k,7}^* & x_{k,6}^* \\
			x_{k+1,2} & x_{k+1,3} & x_{k+1,4} & x_{k,7}^* & x_{k,6}^* & x_{k,5}^* \\
			\vdots  & \vdots  &  \vdots &   \vdots   &    \vdots  &   \vdots \\
			x_{k+1,6} & x_{k+1,7} & x_{k+1,8} & x_{k,3}^* & x_{k,2}^* & x_{k,1}^* \\
		\end{bmatrix}. 
		\label{eq:esprit:Xk+1}
	\end{equation}
	
	With Eq.~(\ref{eq:signalmodel})(\ref{eq:esprit:Xk})(\ref{eq:esprit:Xk+1}), we know that $A$ is a 6 $\times$ 6 steering matrix, 
	and we can derive $\bm{\Gamma}$ to show that the decomposition of $X_k$ and $X_{k+1}$ exists. Mathematically, we have $\bm{\Gamma} = \begin{bmatrix} \bm{\Gamma}_{k} \quad \bm{\Gamma}_{k+1} \end{bmatrix}$, and
	\begin{align}
		\bm{\Gamma}_{k} &= 
		\begin{bmatrix}
			\gamma_1 & \gamma_1\Omega_1 & \gamma_1(\Omega_1)^2 &   \\
			\gamma_2 & \gamma_2\Omega_2 & \gamma_2(\Omega_2)^2&  \\
			\vdots & \vdots & \vdots  \\
			\gamma_P & \gamma_P\Omega_P & \gamma_P(\Omega_P)^2&  \\
		\end{bmatrix}, \label{eq:exprit:gamma:k} \\
		\bm{\Gamma}_{k+1} &= 
		\begin{bmatrix}
			\gamma_1^*(\Omega_1^*)^7 \Phi_1^* & \gamma_1^* (\Omega_1^*)^6\Phi_1^* & \gamma_1^* (\Omega_1^*)^5\Phi_1^*\\
			\gamma_2^*(\Omega_2^*)^7 \Phi_2^* & \gamma_2^* (\Omega_2^*)^6 \Phi_2^* & \gamma_2^* (\Omega_2^*)^5\Phi_2^*\\
			\vdots & \vdots & \vdots  \\
			\gamma_P^*(\Omega_P^*)^7 \Phi_P^* & \gamma_P^* (\Omega_P^*)^6 \Phi_P^*  & \gamma_P^* (\Omega_P^*)^5\Phi_P^*\\
		\end{bmatrix}. \label{eq:exprit:gamma:k+1}
	\end{align}
	
	The rows of $\bm{\Gamma}$ represent multipath signals. Normally, multipath signals (defined in Eq.~(\ref{eq:model:gamma})) are highly correlated. With Eq.~(\ref{eq:exprit:gamma:k}) and (\ref{eq:exprit:gamma:k+1}), the signals are no longer correlated.
	As shown in Eq.~(\ref{eq:esprit:Xk}) and (\ref{eq:esprit:Xk+1}), the number of the measurements is doubled from that without conjugates. Therefore, the rank of covariance matrix is increased from four to six to resolve more possible multipaths. 
	Generally, given the measurements of $M$ channels, the multipaths that conjugated ESPRIT can solve is $\lfloor\frac{2(M+1)}{3}\rfloor$. Our evaluation shows that the proposed algorithm can increase the accuracy of LoRaWAN indoor localization.

	\subsection{Multiple AP fusion}
	\label{subsec:fusion}
	
	\begin{figure}[b]	
		\centering
		\includegraphics[width=0.70\linewidth]{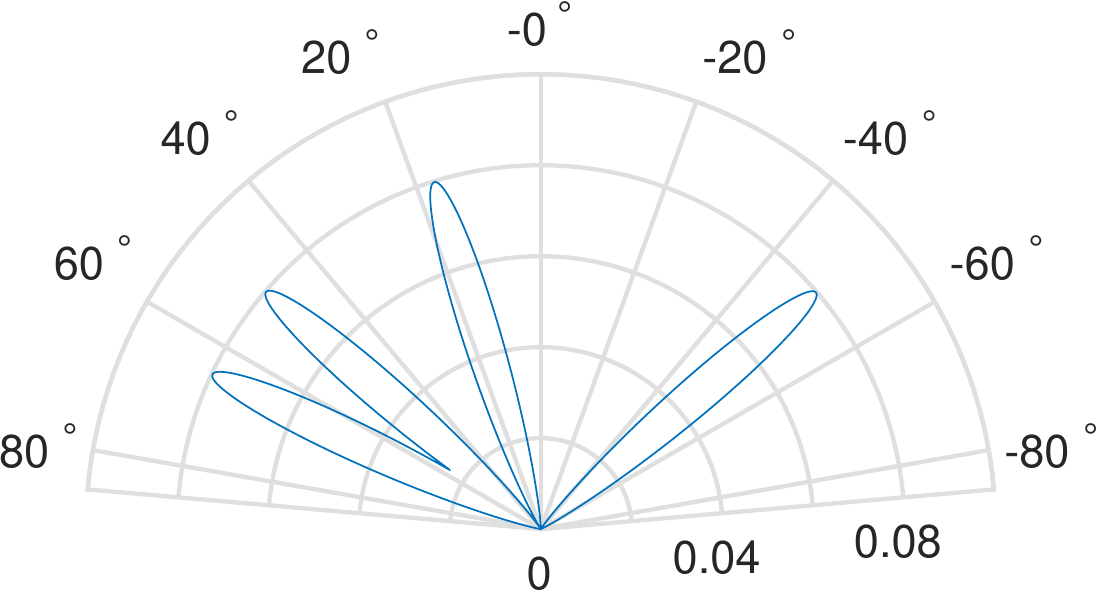}
		\vspace{6pt}
		\caption{An example of a likelihood function with $\sigma=5$. This figure shows four significant paths. The likelihood function can be translated into a heat map, as in Fig.~\ref{fig:heatmap:ap:1}, given the locations of the APs.}
		\label{fig:likelihood}
	\end{figure}
	
	\begin{figure}[htb]
		\centering
		\begin{subfigure}[t]{0.45\linewidth}
			\centering
			\includegraphics[width=\textwidth]{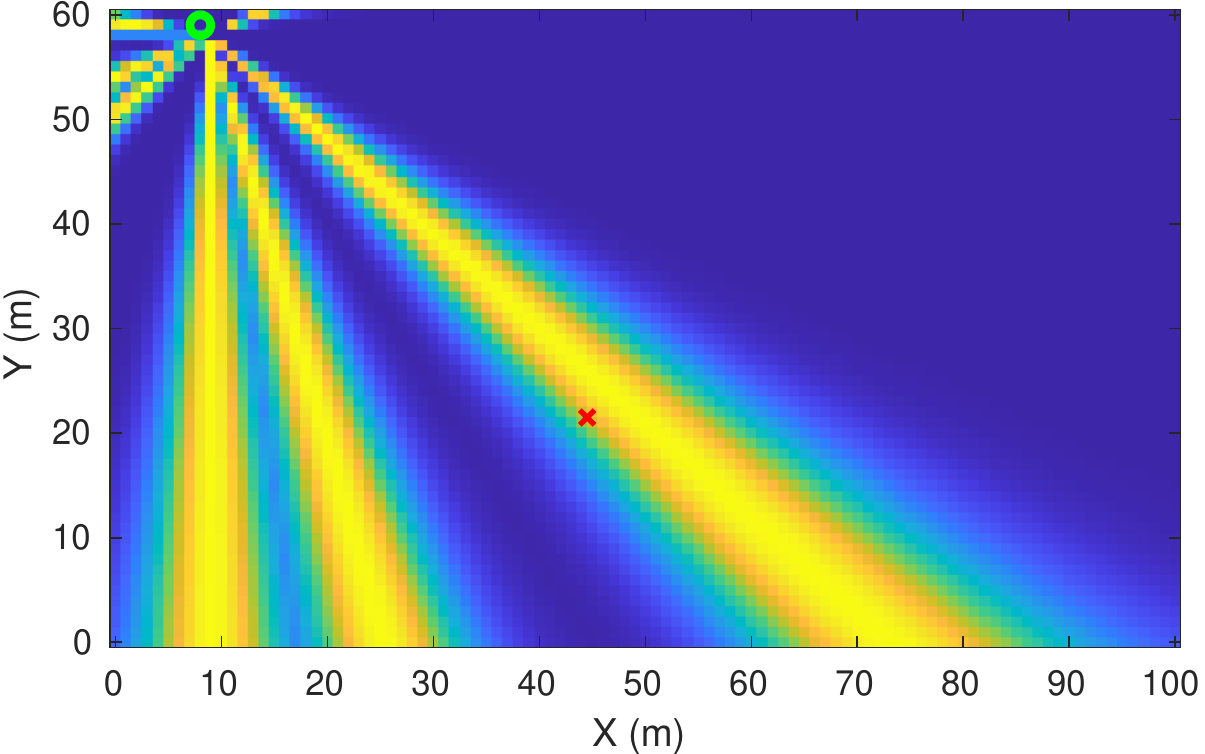}
			\caption{}
			\label{fig:heatmap:ap:1}
		\end{subfigure}
		\quad
		\begin{subfigure}[t]{0.45\linewidth}
			\centering
			\includegraphics[width=\textwidth]{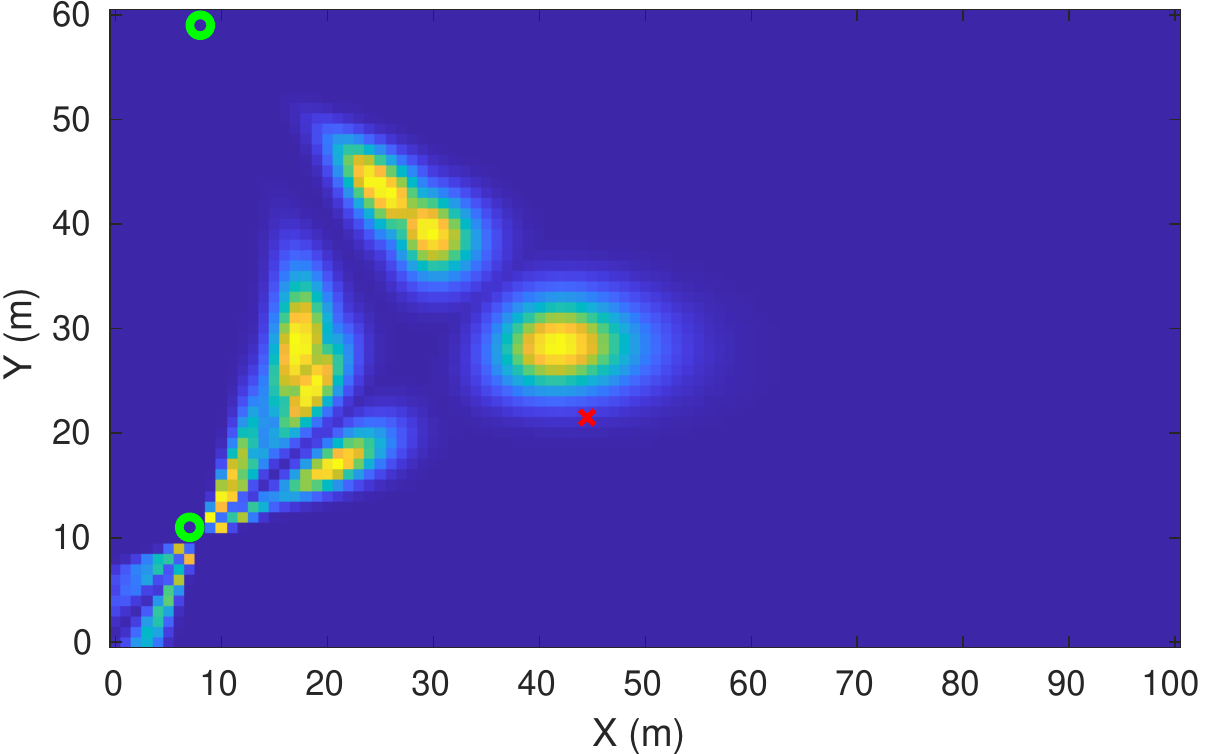}
			\caption{}
			\label{fig:heatmap:ap:2}
		\end{subfigure}
		\hfill
		\begin{subfigure}[t]{0.45\linewidth}
			\centering
			\includegraphics[width=\textwidth]{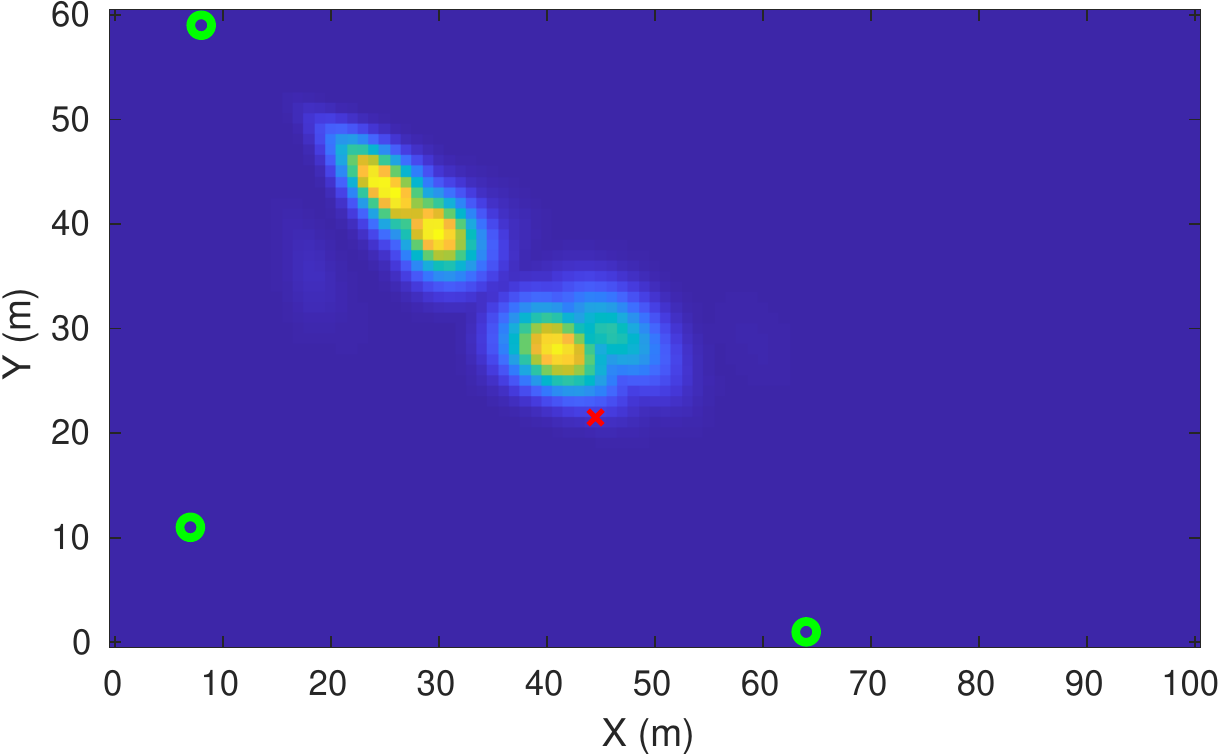}
			\caption{}
			\label{fig:heatmap:ap:3}
		\end{subfigure}
		\quad
		\begin{subfigure}[t]{0.45\linewidth}
			\centering
			\includegraphics[width=\textwidth]{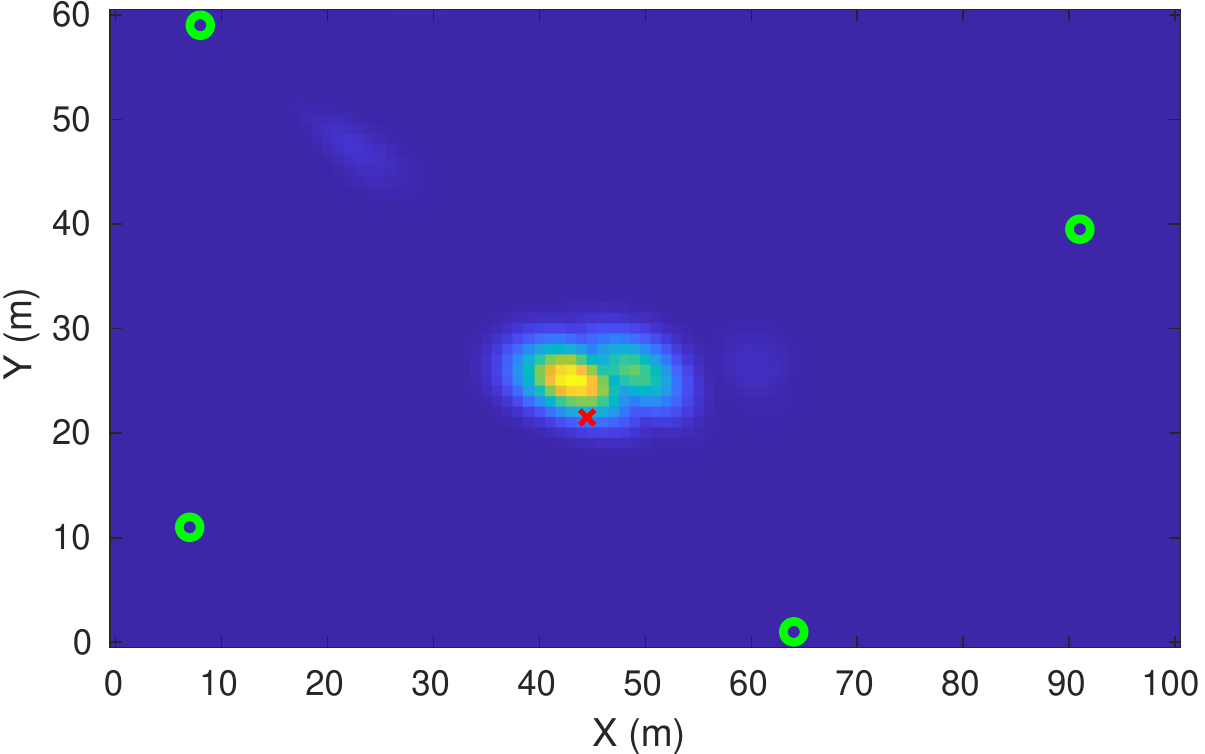}
			\caption{}
			\label{fig:heatmap:ap:4}
		\end{subfigure}
		\caption{A heat map example for multiple AP fusion. Green circles indicate the APs, and the red cross indicates the ground-truth of the transmitter. (a) One AP; multiple AoAs exist. (b) Two APs; seven clusters exist after fusion. (c) Three APs; four clusters exist after fusion. (d) Four APs; two clusters exist. The cluster with stronger likelihood is the location estimation for the transmitter, which is very close to the ground truth. Note: Heat maps view best in color.}
		\label{fig:heatmap:aps}
	\end{figure}

	Previous research proposes determining the direct path based on ToFs~\cite{Kotaru2015,Xiong2015,Vasisht2016}. Similarly, by using AoA-ToF joint estimation (Sec.~\ref{section:tof}), we can determine the direct path and apply triangulation for localization. However, the results of AoA-ToF joint estimation are not reliable, as discussed in Sec.~\ref{section:tof}. Evaluation also suggests that AoA-ToF joint estimation has poor performance in localization (see Sec.~\ref{sec:evaluation}). Alternatively, \textit{Seirios} can exploit the conjugated ESPRIT for reliable AoA estimation (Sec.~\ref{section:esprit}). However, the algorithm proposes all the possible directions of incoming paths regardless of direct paths or reflectors, making it difficult to determine the direct path with one AP only.  
	Therefore, \textit{Seirios} does not resolve the direct path at the beginning. Instead, \textit{Seirios} fusions the AoA estimation of multiple APs to locate the transmitter. This method is based on the fact that the origins of the direct paths are congregated at the transmitter, but that of the reflectors are diverged (see examples in Figs.~\ref{fig:architecture} and~\ref{fig:heatmap:aps}). Different to the classical triangulation algorithm that minimizes the 2-norm of localization errors, \textit{Seirios} utilizes a maximum likelihood algorithm to simplify the calculation.

	Notably, \textit{Seirios} deals with all estimated AoAs equally, as they can be either the direct paths or the reflectors. The errors of AoA estimation are modeled as Gaussian distributions $\mathcal{N}(0,\sigma^2)$. Therefore, we can apply a function $\ell(\theta)$ with the Gaussian distribution's probability density function (PDF) $f(\theta|\mu,\sigma^2)$ to represent the likelihood of a correct estimation for any angle $\theta$. $l(\theta)$ is defined as
	\begin{equation}
		\ell(\theta) = \max_{i=1 \dots P}{f(\theta|\hat{\theta}_i,\sigma^2)} , \quad -85^{\circ}< \theta < 85^{\circ}.
	\end{equation}
	
	where $\hat{\theta}_i$ is the AoA estimation of the $i$th path, and $\sigma$ is determined by the noise level and is a tunable parameter of \textit{Seirios}. 
	We found that the ``good'' values of $\sigma$ are between $3^{\circ}$ and $5^{\circ}$ empirically. Fig.~\ref{fig:likelihood} is an example of the likelihood function $\ell(\theta)$.

	Since the position of APs and the directions of antennas are known, we can translate the likelihood function into a heat map $\mathcal{L}(x,y)$ demonstrating the likelihood of the transmitter's location, as shown in Fig.~\ref{fig:heatmap:ap:1}. As the size of the antenna array is relatively small compared to the distance between the transmitter and the receiver, instead of using hyperbolas to translate the AoAs into a heat map, we use straight lines to simplify the calculation. To fusion the likelihood estimated by multiple APs, \textit{Seirios} merges the heat maps through multiplication. Therefore, the hea tmap $\mathcal{\hat{L}}(x,y)$ as the fusion of $G$ APs is generated by
	\begin{equation}
		\mathcal{\hat{L}}(x,y) = \prod_{g=1}^{G} \mathcal{L}_g(x,y).
	\end{equation}
	
	\begin{figure*}[ht]
		\centering
		\includegraphics[width=0.84\linewidth]{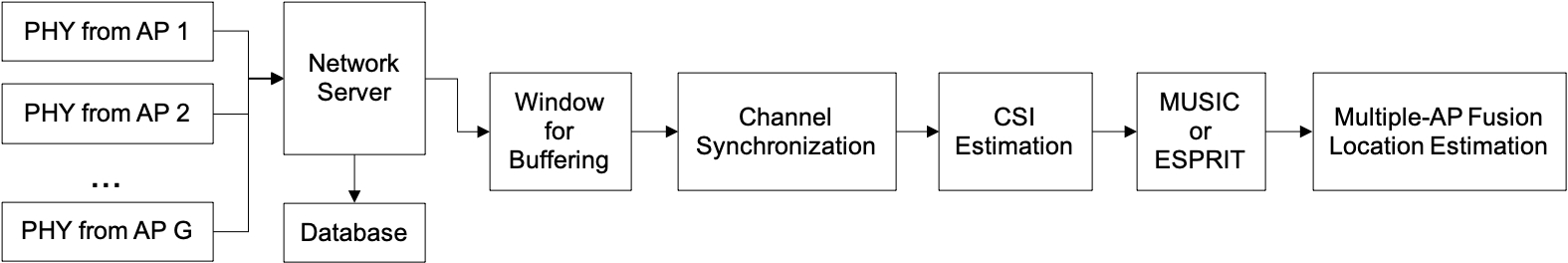}
		\caption{Flowchart for signal processing in the \textit{Seirios} cloud service.}
		\label{fig:server:flowchart}
		
		CSI: channel state information; PHY: physical layer 
	\end{figure*}

	However, there might be multiple clusters if the number of APs is insufficient. 
	Increasing the APs can subsequently reduce ambiguity.
	Fig.~\ref{fig:heatmap:aps} shows the refining of location estimation as the number of APs increases. On the heat map with two APs (Fig.~\ref{fig:heatmap:ap:2}), there are many possible locations for the transmitter. With four APs (see Fig.~\ref{fig:heatmap:ap:4}), the transmitter's location is estimated as the center of a cluster, which is very close to the ground truth (i.e., the red cross). With the multiple AP fusion algorithm, it is unnecessary to determine the direct path for each AP, which can avoid inaccurate ToF estimation, as discussed in Sec.~\ref{section:tof}. 
	
	LoRa signals are good at penetration, hence, why in most of the cases there exist direct paths. The study and evaluation of cases in which direct paths are completely blocked can be addressed in future studies.

	\section{Implementation}
	\label{sec:implementation}

	\begin{figure}[b]
		\centering
		\begin{subfigure}[t]{0.35\linewidth}
			\centering
			\includegraphics[width=\textwidth]{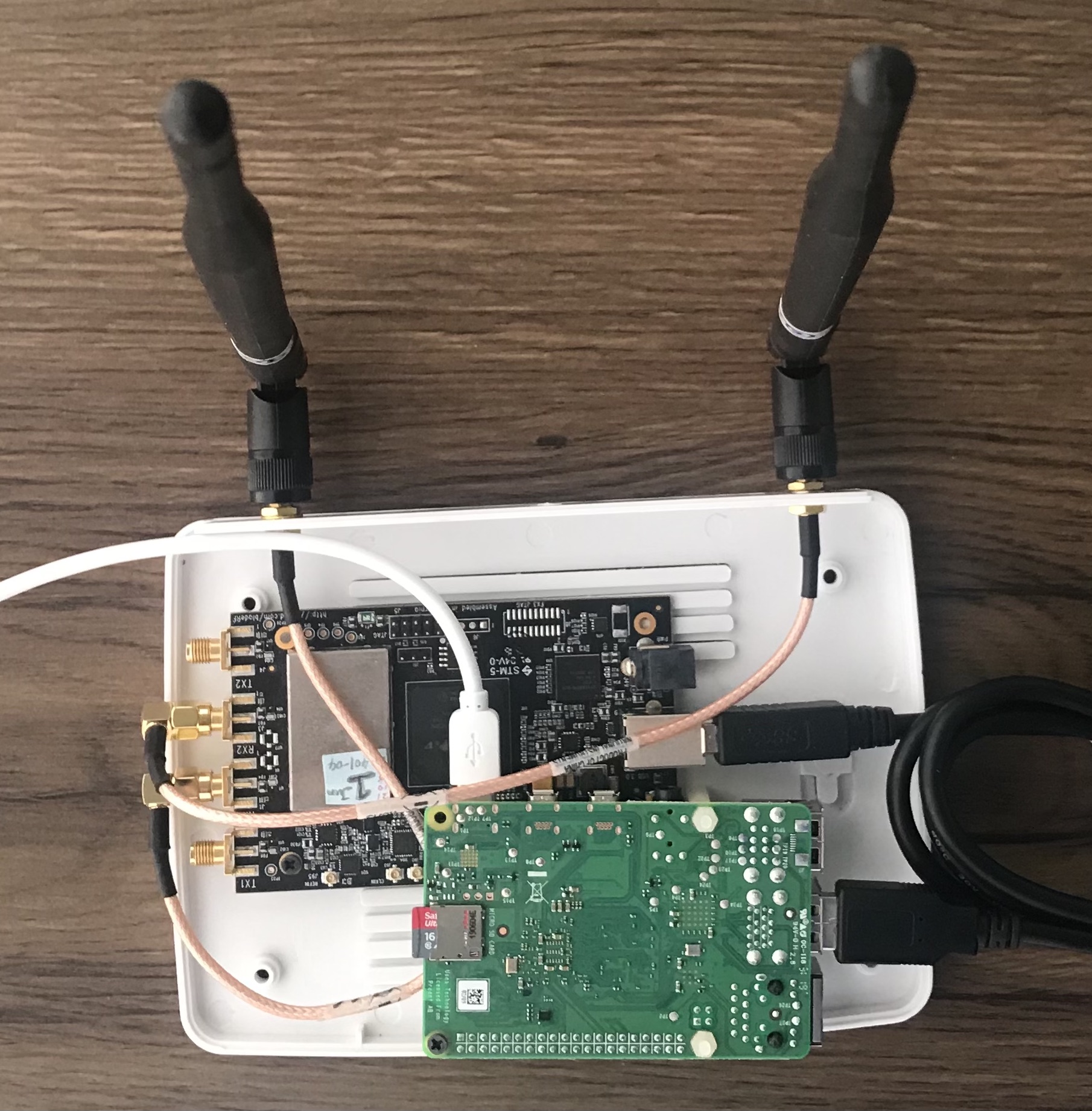}
			\caption{}
			\label{fig:implementation:ap}
		\end{subfigure}
		\qquad
		\begin{subfigure}[t]{0.35\linewidth}
			\centering
			\includegraphics[width=\textwidth]{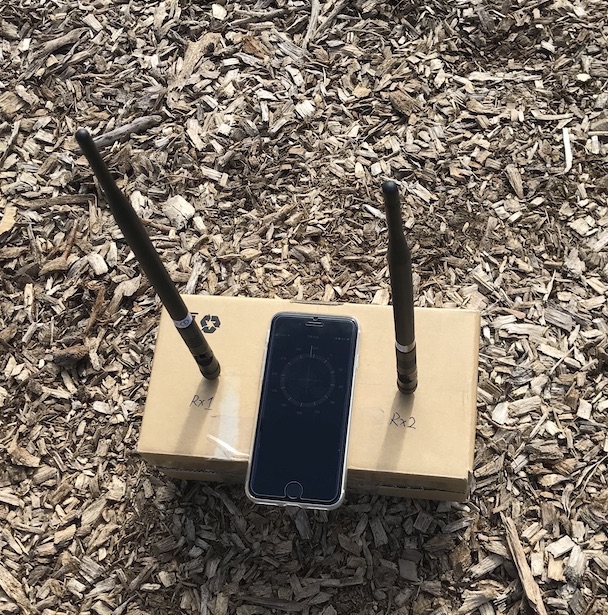}
			\caption{}
			\label{fig:implementation:apcase}
		\end{subfigure}
		\caption{(a) AP implementation for data acquisition. (b) AP for outdoor deployment.}
	\end{figure}

	\textit{Seirios} is designed for transmitter localization in an outdoor or uncluttered indoor environment. According to the architecture shown in Fig.~\ref{fig:architecture}, we implemented a \textit{Seirios}
	prototype with the APs for wireless data acquisition and the cloud service for data processing. 
	
	\textit{Seirios AP prototype}. We used bladeRF 2.0 SDR, which supports 2 $\times$ 2 MIMO as the AP prototype to receive LoRa radio signals between 902 MHz and 928 MHz. The SDR has two antennas (LoRa gateways usually have only two antennas~\cite{xie2020combating,zhang2020exploring}), and can generate two synchronized $I$ and $Q$ streams with 12-bit resolution. The distance between antennas is fixed at 14cm, which is slightly less than half of the radio wavelength. Prior calibration is performed to eliminate the phase offset caused by connectors or cables. The SDR is connected to a general-purpose processor (GPP) via USB 3.0, which can be either a PC or a single board
	embedded computer. To operate the \textit{Seirios} prototype in a mobile manner (e.g., in an outdoor environment),  we decided to use a low-power embedded system Raspberry Pi 4 as the GPP, which is deployed with a signal processing program for LoRa packet detection. The signal processing algorithm is implemented in C++ with GNU Radio for high efficiency. Once a LoRa packet is detected, the GPP uploads the packet to the \textit{Seirios} cloud service
	prototype using high-speed WiFi. The devices are shown in Fig.~\ref{fig:implementation:ap}. For outdoor deployment, the devices are packed in a case (see Fig.~\ref{fig:implementation:apcase}).

	\textit{Cloud server}. The cloud has a TCP server for incoming LoRa PHY that is uploaded by multiple APs. The server stores the data with a timestamp and maintains a time window to group the relevant packets for further processing. For each AP, when at least one packet for each channel is recorded in the window, the server will start to process the data. It first extracts and synchronizes the CSI for each channel. Then, it performs either MUSIC or ESPRIT for the AoA estimation. After that, the server will perform multiple-AP fusion to estimate the location of the transmitter. A flowchart of the process is shown in Fig.~\ref{fig:server:flowchart}.

	\section{Microbenchmark}
	\label{sec:microbenchmark}

	\begin{figure*}[htb]
		\centering
		\begin{subfigure}{0.31\textwidth}
			\centering
			\includegraphics[width=\linewidth]{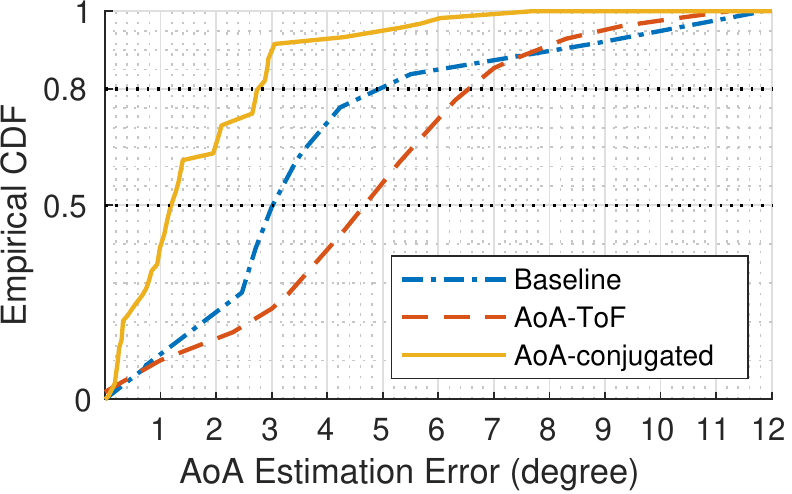}
			\caption{Line-of-sight (LoS)}
			\label{fig:micro:aoa:los}
		\end{subfigure}  
		\hfill
		\begin{subfigure}{0.31\textwidth}
			\centering
			\includegraphics[width=\linewidth]{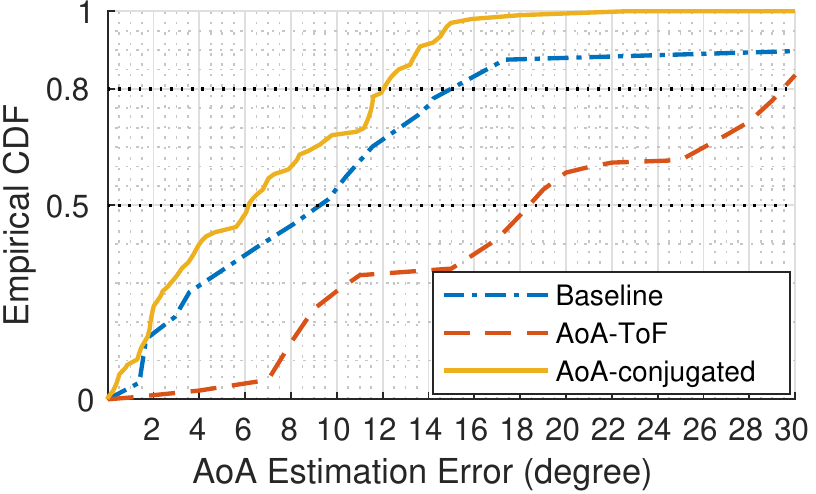}
			\caption{Non-line-of-sight (NLoS)}
			\label{fig:micro:aoa:nlos}
		\end{subfigure}  
		\hfill
		\begin{subfigure}{0.31\textwidth}
			\centering
			\includegraphics[width=\linewidth]{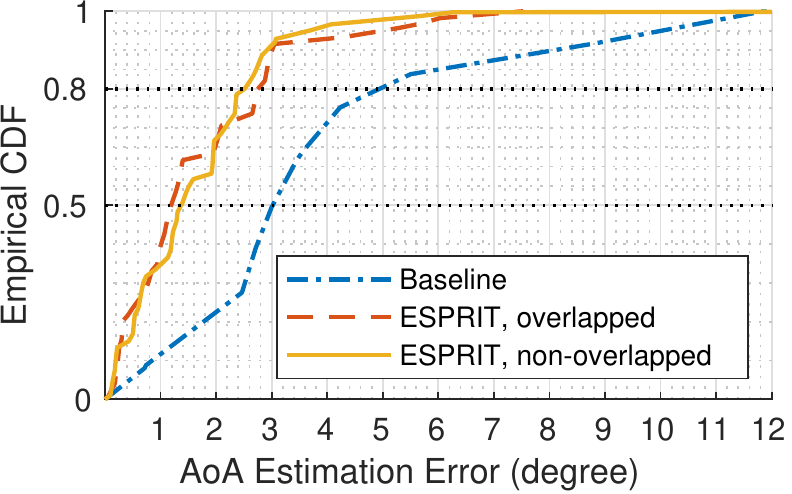}
			\caption{Overlapped and non-overlapped}
			\label{fig:micro:interchannelsync}
		\end{subfigure}  
		\caption{Microbenchmark: (a) LoS, (b) NLoS, and (c) overlapped and non-overlapped channels. The accuracy of the two cases is close, with half the error as the baseline. The performance of the interchannel synchronization algorithm for non-overlapped channels is comparable to that for overlapped channels. }
		\raggedright AoA: angle of arrival; CDF: cumulative distribution function; ToF: time of flight
	\end{figure*}

	So far, we have discussed interchannel synchronization  (Sec. \ref{section:sync}) and conjugated ESPRIT (Sec.~\ref{section:esprit}) as the key algorithms of \textit{Seirios} to improve the accuracy of AoA estimation as well as the localization of LoRaWAN IoT devices. To understand the performance of the algorithms on one AP, we conducted a microbenchmark indoors (25 m $\times$ 15 m) with LoS and non-line of sight (NLoS) to evaluate the AoA estimation accuracy. We compared three algorithms (i.e. AoA-conjugated using conjugated ESPRIT, Sec.~\ref{section:esprit}; AoA-ToF joint estimation using MUSIC, Sec.~\ref{section:tof}; and the baseline). The baseline algorithm is similar to TDoA, but instead of synchronizing multiple APs for timestamp measurement, we synchronized two antennas of one AP to extract the time difference based on the phase difference and further calculated AoA based on the measurements. 
	
	\subsection{AoA Estimation Accuracy}
	\label{sec:microbenchmark:aoa}
	
	Figs.~\ref{fig:micro:aoa:los} and \ref{fig:micro:aoa:nlos} show the cumulative distribution function (CDF) for AoA estimation errors of the AoA-conjugated, AoA-ToF, and the baseline, for LoS and NLoS, respectively. The lab where the data were collected is a typical cluttered indoor environment. For LoS, AoA-conjugated (median error $1.2^\circ$) has three times superior accuracy compared to AoA-ToF ($4.7^\circ$) and 1.5 times compared to the baseline ($3.0^\circ$). Compared to the previous observation in a Wi-Fi localization system, AoA-ToF joint estimation has even worse performance than the baseline (see Sec.~\ref{section:tof} for an explanation of this result). For NLoS, the radio path is completely blocked by walls so that only penetrated radio (with much noise) can reach the receiver. The median error ($6.0^\circ$, $18.2^\circ$, and $9.2^\circ$ for AoA-conjugated, AoA-ToF and the baseline, respectively) is larger than those under LoS, but AoA-conjugated still has the best performance. Therefore, we can conclude that AoA estimation with conjugated ESPRIT can, on average, improve the accuracy by two times compared to the baseline.  
	
	\subsection{Interchannel Synchronization}
	\label{section:evaluation:sync}

	If using customized LoRa protocol~\cite{8710297} instead of LoRaWAN, one may define overlapped LoRa channels, and exploit them to improve localization accuracy.
	However, the communication channels defined by LoRaWAN do not overlap (to avoid interchannel interference and improve transmission performance); therefore, a virtual intermediate channel must be synthesized to synchronize LoRaWAN channels, as discussed in Sec.~\ref{section:sync}. Theoretically, synchronization for non-overlapped channels may introduce extra errors compared to overlapped channels, which may increase the AoA estimation and localization errors. In practice, the extra error is relatively small.
	
	For demonstration, we conducted a microbenchmark to compare the
	performance of \textit{Seirios} in non-overlapped and overlapped LoRa channels. 
	To have overlapped channels, a LoRa transmitter is programmed to transmit packets in 15 channels of 125 kHz with the channel spacing of 100 kHz, 30 times over. We selected odd numbers of channels (i.e., $1$st, $3$rd, $\dots$ $15$th) from the overlapped dataset to form a ``new'' dataset of non-overlapped channels.

	Then, we performed conjugated ESPRIT on both datasets for AoA estimation. Fig.~\ref{fig:micro:interchannelsync} shows that there is insignificant difference between the two datasets (i.e., overlapped and non-overlapped). Taking the median error for comparison, the loss ($0.2^\circ$) of the non-overlapped dataset is very small and is only one-eighth of its improvement from the baseline. Therefore, the error introduced by non-overlapped channels has little effect on AoA estimation.

	\section{Evaluation}
	\label{sec:evaluation}

	\subsection{Goals, Metrics and Methodology}
	\label{subsec:goals}
\begin{figure}[bt]
		\centering
		\includegraphics[width=0.60\linewidth]{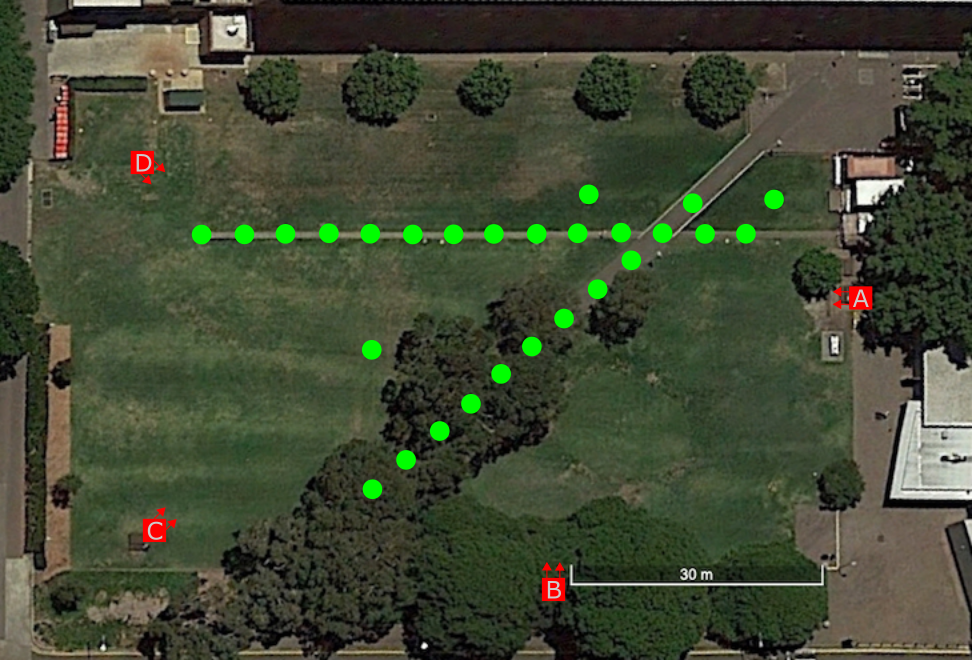}
		\caption{Outdoor evaluation on 100 m $\times$ 60 m campus lawn. The red squares marked with A/B/C/D indicate the APs. The green circles indicate the ground truth of transmitter locations. 
		}
		
		\label{fig:outdoormap}
	\end{figure}

\begin{figure}[htb]
		\centering
		\vspace{-6pt}
		\includegraphics[width=0.6\linewidth]{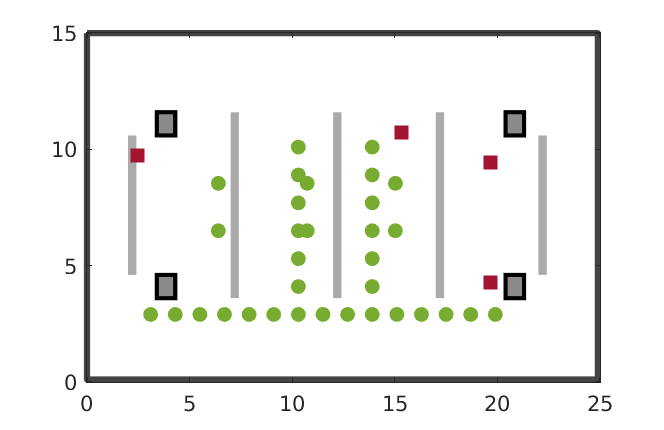}
		\caption{Indoor evaluation on a 25 m$\times$ 15 m large room. The red squares indicate the APs, and the green circles indicate the ground truth of transmitter locations. The four black and gray rectangles are concrete pillars that may cause strong reflection. The gray bars are 1.5 m tall barriers used to split the lab into several zones, which blocked the LoS paths. 
		}
		\raggedright Note: Values are measured in meters.
		
		\label{fig:indoormap}
	\end{figure}

	Our goal in this evaluation was to show that \textit{Seirios} can locate a LoRa 
	transmitter accurately for both outdoor and indoor environments. For this purpose, we evaluated the performance 
	of the \textit{Seirios} prototype developed in Sec.~\ref{sec:implementation} in 
	a 100 m $\times$ 60 m lawn with several trees, surrounded by buildings (see Fig.~\ref{fig:outdoormap}), and in a 25 m$\times$ 15 m large room with concrete pillars, surrounded by walls (see Fig.~\ref{fig:indoormap}). 
	
	For outdoor scenarios, we deployed four APs on the lawn, one of which is shown in Fig.~\ref{fig:implementation:apcase}. A pair of 5 dBi antennas was mounted on each AP. The devices were battery powered and each were sampling at 2 MSps to cover
	the 1.6 MHz LoRaWAN spectrum. We used mDots\footnote{
		Hyperlink will be revealed in the final version.
	} as the transmitters (i.e.,
	the embedded LoRaWAN devices to be localized), which were configured with frequency hopping at eight
	LoRaWAN channels. The APs can overhear the packets sent by the transmitters on each channel and use them for localization. For convenience, the transmitters were configured with burst transmission, and each of them transmitted packets for $SF=7$ from channel 0 to channel 7, with the channel spacing of 200 kHz. It took a device approximately 30 ms to send a predefined packet; therefore, it took approximately $30 \times 8 = 240$ ms to cover all the channels. The transmit power was fixed as 14 dBm.
	
	During the evaluation, each of the transmitters transmitted LoRaWAN packets in all channels a total of three times. At the same time, we recorded the locations of the transmitters as the ground truth (see the
	green dots and the red squares in Fig.~\ref{fig:outdoormap} for the locations
	of the transmitters and APs, respectively). 
	
	Similarly, we deployed APs and transmitters in our lab for indoor evaluation. As mentioned, the lab is a cluttered environment with furniture and walls (see Fig.~\ref{fig:indoormap}).
	The transmitters were configured similarly to those for outdoor evaluation.
	
	The metrics used to evaluate the performance of \textit{Seirios} is 
	the error of localization (in meters), which is simply the absolute distance between the estimation and the ground truth. 
	
	Besides overall localization performance with conjugated ESPRIT (Sec.~\ref{section:esprit}), AoA-ToF joint estimation (Sec.~\ref{section:tof}), and the baseline (Sec.~\ref{sec:microbenchmark}), we also investigated the performance of different components of \textit{Seirios} such as 
	the fusion algorithm (see Sec.~\ref{subsec:fusion}), and the effectiveness of conjugated ESPRIT.

	\subsection{Outdoor Localization}
	\label{subsec:out}

	\begin{figure*}[htb]
		\centering
		\begin{subfigure}{0.32\textwidth}
			\centering
			\includegraphics[width=\linewidth]{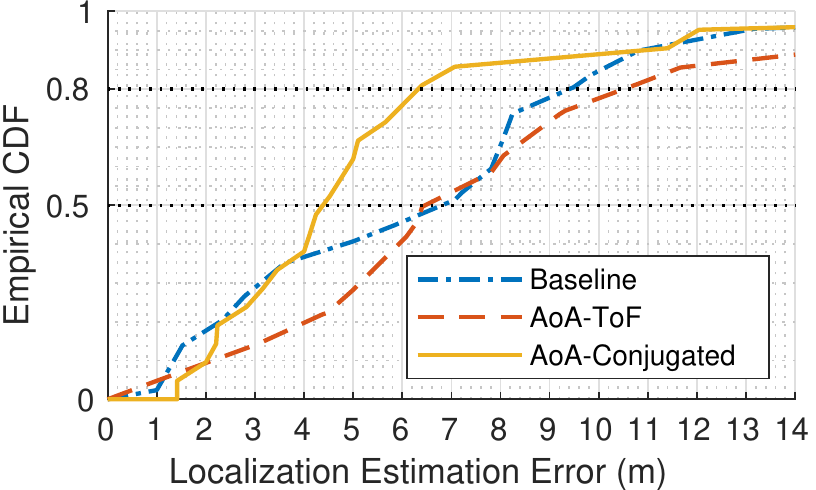}
			\caption{Outdoor localization error}
			\label{fig:cdf:localization:outdoor}
		\end{subfigure}    
		\hfill
		\begin{subfigure}{0.32\textwidth}
			\centering
			\includegraphics[width=\linewidth]{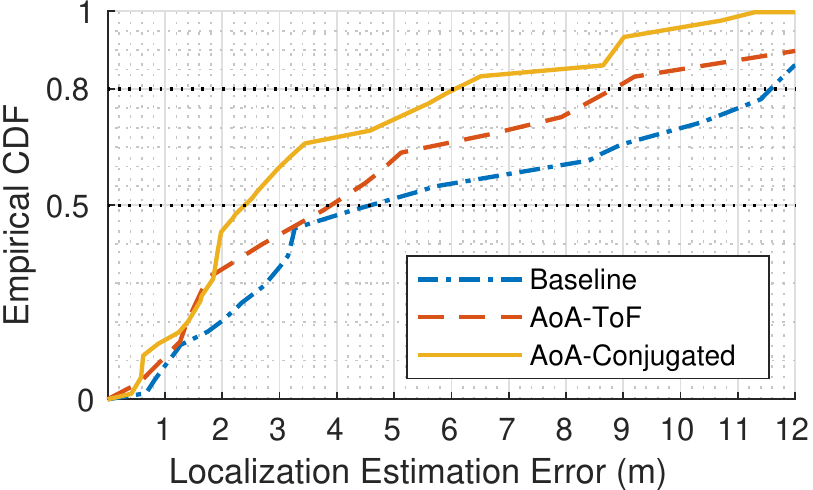}
			\caption{Indoor localization error}
			\label{fig:cdf:localization:indoor}
		\end{subfigure}
		\hfill
		\begin{subfigure}{0.32\textwidth}
			\centering
			\includegraphics[width=\linewidth]{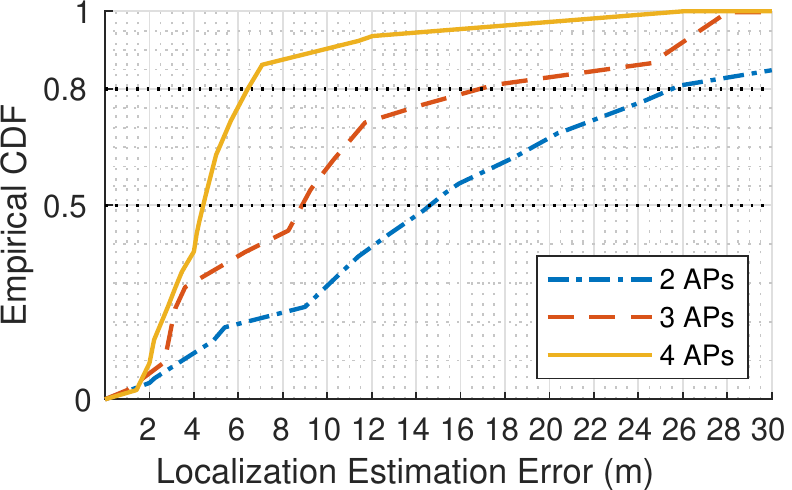}
			\caption{Effect of APs evaluated outdoors}
			\label{fig:cdf:numberofap:out}
		\end{subfigure}    
		\begin{subfigure}{0.32\textwidth}
			\centering
			\includegraphics[width=\linewidth]{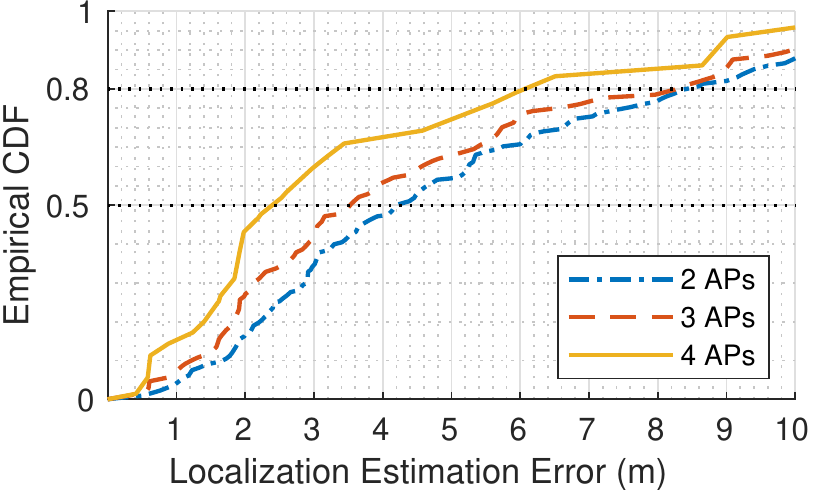}
			\caption{Effect of APs evaluated indoors}
			\label{fig:cdf:numberofap:in}
		\end{subfigure}    
		\hfill
		\begin{subfigure}{0.32\textwidth}
			\centering
			\includegraphics[width=\linewidth]{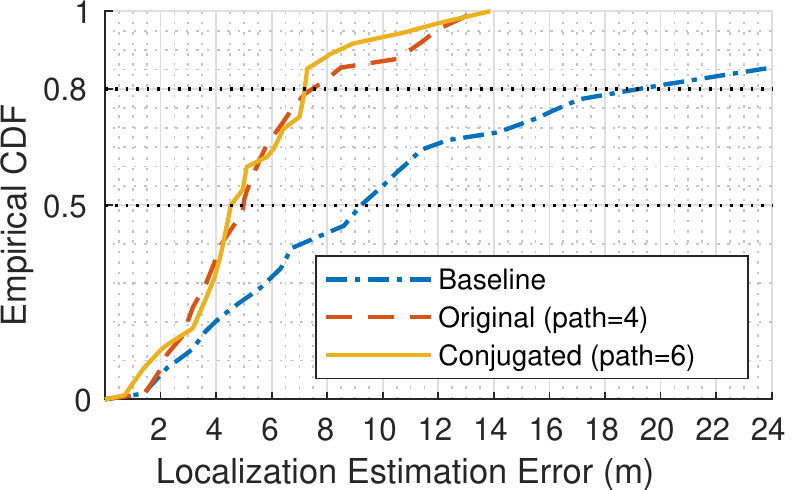}
			\caption{Multipath effect evaluated outdoors}
			\label{fig:cdf:numberofmultipath:out}
		\end{subfigure}
		\hfill
		\begin{subfigure}{0.32\textwidth}
			\centering
			\includegraphics[width=\linewidth]{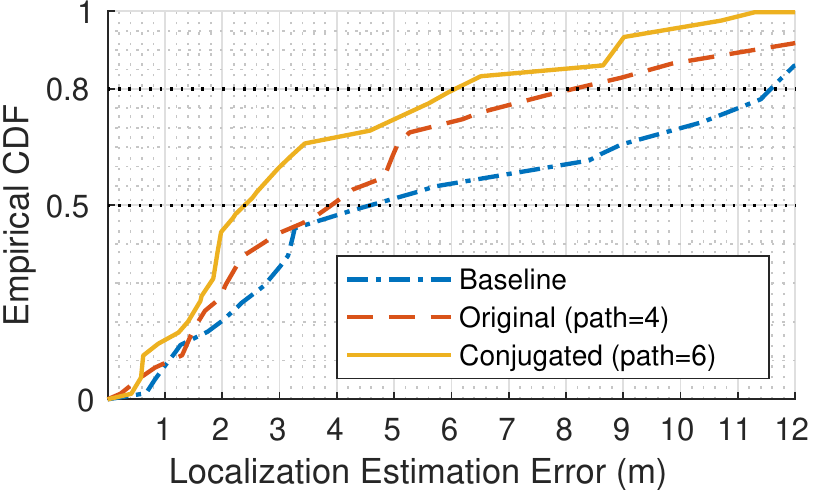}
			\caption{Multipath effect evaluated indoors}
			\label{fig:cdf:numberofmultipath:in}
		\end{subfigure}    
		\caption{(a) Outdoor localization error. (b) Indoor localization error. (c) Effect of APs evaluated outdoors. (d) Effect of APs evaluated indoors. (e) The effect of multipath evaluated outdoors. (f) The effect of multipath evaluated indoors.}
	\end{figure*}

	For the outdoor evaluation on the campus lawn shown in Fig.~\ref{fig:outdoormap}, the overall performance of \textit{Seirios} is shown in Fig.~\ref{fig:cdf:localization:outdoor}. The performance of conjugated ESPRIT is
	better than that of AoA-ToF joint estimation and the baseline with the median localization errors of 4.4 m, 6.4 m, and 6.9 m for conjugated ESPRIT, AoA-ToF joint estimation, and the baseline, respectively. The 80th percentiles are 6.4 m, 10.5 m, and 9.4 m for
	conjugated ESPRIT, AoA-ToF joint estimation, and the baseline, respectively. 
	
	The results show that the localization error is reduced by 36.2\% when comparing conjugated ESPRIT with the baseline. Further,
	\textbf{conjugated ESPRIT has better performance than AoA-ToF joint estimation, which is different to the results reported in previous Wi-Fi localization literature}. This phenomenon is due to the foundation of the model discussed in Sec.~\ref{section:tof} that an accurate AoA estimation with AoA-ToF joint estimation relies on an accurate ToF estimation, while the ToF estimation is sensitive to the raw resolution limited by the overall bandwidth of LoRaWAN channels that is orders of magnitude smaller than that of Wi-Fi channels. AoA-ToF joint estimation performs worse than the baseline, which shows that the error introduced by an inaccurate ToF estimation reduce the performance of AoA estimation.
	
	The results also indicate that super-resolution algorithms are useful for outdoor environments. Even though the number of multipaths is smaller than that of indoor environments, the strong radio signal reflection can be caused by trees and nearby buildings, which increase localization errors.

	\subsection{Indoor Localization}
	\label{subsec:in}
	\label{subsec:in:localization_accuracy}

	For the indoor evaluation in our lab, as shown in Fig.~\ref{fig:indoormap}, the overall performance of \textit{Seirios} is shown in Fig.~\ref{fig:cdf:localization:indoor}. The median errors with conjugated ESPRIT, AoA-ToF joint estimation, and the baseline are 2.4 m, 4.0 m, and 4.6 m, respectively, and the 80th percentiles are 6.1 m, 8.8 m, and 11.6 m, respectively. The results show that with the conjugated ESPRIT, the localization error is reduced by 47.8\% compared to the baseline, which is slightly more than that of outdoor evaluation (36.2\%, see Sec.~\ref{subsec:out}). This is because the multipath effect is more severe indoors than outdoors, which affects the baseline algorithm but can be resolved by the conjugated ESPRIT. On average (indoors and outdoors), the localization improvement is 42\%, which demonstrates the superior localization performance of \textit{Seirios} (i.e., conjugated ESPRIT) on narrow bandwidth radio signals with commercially available off-the-shelf (COTS) hardware, which has
	two antennas only.
	
	Unlike outdoor evaluation, AoA-ToF joint estimation is better than the baseline indoors. The reason behind this phenomenon is that ToF estimation is apt to average the ToF of multipaths if they are not distinguishable, and indoor paths have relatively similar ToF, so the estimation is relatively more accurate than those of outdoors evaluations (with significantly larger different ToF for each path). The relatively accurate ToF estimation can improve the accuracy of AoA estimation, so that the localization performance of AoA-ToF joint estimation is better than the baseline. Nevertheless, multipaths are not resolvable with AoA-ToF joint estimation, so its performance is worse than that of conjugated ESPRIT.
	
	\subsection{Comparison with RSS and TDoA-based Approaches}
	
	As discussed in Sec.~\ref{section:introduction}, RSS and TDoA based LoRaWAN localization systems have poor performance. In this section, we evaluate these techniques in both indoor and outdoor environments as a comparison to \textit{Seirios}.
	For the RSS-based approach, we follow a general path-loss model proposed in \cite{liu2007survey,schmid2010approach} with triangulation for evaluation. For the TDoA-based approach, we use the phase difference of two synchronized antennas in each gateway to calculate TDoA. 
	
	The median \textbf{outdoor} localization errors are 15.3 m and 6.9 m for RSS and TDoA based approaches, respectively, while the median \textbf{indoor} localization errors are 6.3 m and 4.6 m for RSS and TDoA based approaches, respectively. Therefore, the AoA-based conjugated ESPRIT 
	of \textit{Seirios} produces
	significantly better performance (4.4 m outdoors, 2.4 m indoors) than these of RSS and TDoA based approaches.

	\subsection{Impact of AP Fusion}
	\label{sec:evaluation:apnumber}

	AoA estimation proposes the possible directions of all incoming paths regardless of the direct path or reflectors. AP fusion is an effective technique to determine the direct path as well as the location of the transmitter. It is based on the fact that the direct paths are congregated but the reflectors are diverged.
	
	\textit{Seirios} uses multiple-AP fusion for localization. As shown in Fig.~\ref{fig:heatmap:aps}, fusing more APs can reduce the ambiguity of localization estimation and increase its accuracy. To investigate how the number of APs affects the overall performance, we evaluated the localization errors with two, three, and four APs, respectively. 
	
	For outdoor environments, we first evaluated two APs with \textit{C} and \textit{D}, as shown in Fig~\ref{fig:outdoormap}, and then added APs \textit{B} and \textit{A} sequentially for the evaluation of three and four APs, respectively. The results are shown in Fig.~\ref{fig:cdf:numberofap:out}. With two APs, the median and 80th percentile errors are 14.8 m and 26.0 m, respectively. By increasing the number of APs from two to three, the median and 80th percentile errors are reduced to 8.8 m and 17.0 m, respectively. With four APs, the median error is reduced by 50\% of those with three APs and 70\% of those with two APs to 4.4 m, and the 80th percentile error is reduced to 6.4 m.
	
	For indoor environments, we observe similar behavior, as shown in Fig.~\ref{fig:cdf:numberofap:in}. Here, the error reduction with four APs for indoors is 31\% compared to that of three APs and 42\% compared to that of two APs, respectively, which is less than that of outdoor calculations. The reason is that the indoor area is smaller than the outdoor area, but both are equipped with the same number of APs, which means that the indoor area has better LoRaWAN coverage than the outdoor area. Thus, the improvement with extra APs indoors is not as significant as that outdoors.

	Overall, the evaluation shows that localization accuracy relates to the density of APs (gateways) and proves the effectiveness of AP fusion algorithm introduced in Sec.~\ref{subsec:fusion}.

	\subsection{Effectiveness of Conjugates}
	\label{subsec:impactofmultipath}

	Sec.~\ref{section:esprit} proposed to use conjugated ESPRIT to increase the capability of multipath resolution. Theoretically, the proposed algorithm can increase the capacity for multipaths resolution from four to six in our setting (i.e., one AP prototype with two antennas). 
	
	Fig.~\ref{fig:cdf:numberofmultipath:in} shows the localization accuracy achieved by the conventional ESPRIT, the conjugated ESPRIT, and the baseline in the indoor evaluation. The median errors for the aforementioned three algorithms are 2.4 m, 3.9 m and 4.6 m, respectively. Evidently, the conjugated ESPRIT reduced the error significantly. However, Fig.~\ref{fig:cdf:numberofmultipath:out} shows that the conjugated ESPRIT for the outdoor environment does not have as significant an improvement as that of the indoor environment (Fig.~\ref{fig:cdf:numberofmultipath:in}). This is because the outdoor environment has fewer multipaths that can even be solved by the conventional ESPRIT.
	Further, Figs.~\ref{fig:cdf:numberofmultipath:in} and~\ref{fig:cdf:numberofmultipath:out} imply that the number of significant multipaths indoors is more than four, while that for the outdoors is less than or equal to four.

	\section{Limitations and Future Work}
	\label{section:limitation}
	\label{section:discussion}
	\label{section:futurework}

	One limitation of our research is that \textit{Seirios} focuses on one transmitter only and does not support concurrent transmissions from other LoRaWAN devices. However, since LoRa networks benefit hugely from the innate orthogonality of SFs, we believe that concurrent transmissions can be regarded as the noise and suppressed by the super-resolution algorithms~\cite{schmidt1986multiple,roy1986esprit,roy1989esprit}. In this regard, \textit{Seirios} can potentially locate multiple transmitters at the same time. Nevertheless, we leave research in this direction as future work.

	Another limitation is the number of gateways in practice. Normally, one LoRaWAN gateway can cover up to 10 km and thus the density of deployment is low, which is different to our evaluation setup in Sec.\ref{sec:evaluation:apnumber} with multiple gateways for high localization accuracy. However, recent research \cite{Dongare2018,liu2020nephalai} shows that a dense deployment of LoRaWAN gateways improves signal quality, battery lifetime, network scalability and network robustness. We envision the dense deployment of LoRaWAN gateways in the future and leave how the gateway deployment density influences localization performance as future work.
	
	In future work, it will be interesting to investigate if the localization performance of \textit{Seirios} can be further improved by other signal super resolution algorithms, such as compressive sensing and matrix pencil decomposition, and by combining the information from other radio signals such as Wi-Fi and BLE in more cluttered environments.
	\section{Conclusion}
	\label{section:conclusion}
	
	This paper introduced \textit{Seirios}, an AoA-based localization system for LoRaWAN devices. Despite the huge success
	and popularity of AoA-based localization methods in wideband radio systems such as Wi-Fi, no prior studies have explored this method for the emerging narrowband LoRaWAN signal because of the bandwidth limitation that results in poor location estimation. 
	\textit{Seirios}
	addresses this limitation using a novel (non-overlapped) interchannel synchronization method and an ESPRIT algorithm that exploits both the original and the conjugate of the channel state measurements. Our empirical evaluation
	shows that \textit{Seirios} can reduce localization errors by 41.6\% compared to the baseline, and can achieve 4.4 m median accuracy in an open area of 100 m $\times$ 60 m as well as 2.4 m median accuracy in an indoor area of 25 m $\times$ 15 m.

	\bibliographystyle{ACM-Reference-Format}
	\bibliography{localization}
	
\end{document}